\documentclass[5p]{elsarticle}
\usepackage[utf8]{inputenc}
\usepackage{textcomp}
\usepackage[table]{xcolor}
\usepackage{epsfig}
\usepackage{pgfplotstable}
\usepackage{pgfplots}
\usepackage{amsmath}
\usepackage{amssymb}
\usepackage{bbm}
\usepackage{float}
\usepackage{breakurl} 
\usepackage[breaklinks]{hyperref}
\hypersetup{breaklinks=true}

\makeatletter
\providecommand{\doi}[1]{%
  \begingroup
    \let\bibinfo\@secondoftwo
    \urlstyle{rm}%
    \href{http://dx.doi.org/#1}{%
      doi:\discretionary{}{}{}%
      \nolinkurl{#1}%
    }%
  \endgroup
}
\makeatother

\usepackage{microtype}
\usepackage{tikz}
\usetikzlibrary{decorations.text,calc,shapes,arrows,arrows.meta, positioning,shapes.misc,decorations.markings,decorations.markings,decorations.pathreplacing,matrix}
\usepackage{booktabs}
\usepackage{multirow}
\usepackage{adjustbox}
\usepackage{verbatim}
\usepackage[T1]{fontenc}
\usepackage[title]{appendix} %
\usepackage[font={small,singlespacing},labelfont=it,labelsep=period]{caption}
\usepackage{adjustbox} %
\usepackage{graphicx}
\usepackage{caption}
\usepackage{subcaption}
\usepackage[noend]{algpseudocode}
\usepackage[section]{placeins} %
\usepackage{algorithm}
\usepackage{siunitx} %
\usepackage[colorinlistoftodos,disable]{todonotes}

\journal{Digital Investigation}

\usepackage{xspace}

\def\lzjdh{$\text{LZJD}_h$\xspace}
\def\lzjdf{$\text{LZJD}_f$\xspace}

\makeatletter
\pgfplotsset{
    every axis x label/.append style={
        alias=current axis xlabel
    },
    legend pos/outer south/.style={
        /pgfplots/legend style={
            at={%
                (%
                \@ifundefined{pgf@sh@ns@current axis xlabel}%
                {xticklabel cs:0.5}%
                {current axis xlabel.south}%
                )%
            },
            anchor=north
        }
    }
}

\begin{document}

\begin{frontmatter}

\title{Lempel-Ziv Jaccard Distance, an Effective Alternative to Ssdeep and Sdhash}
\tnotetext[t1]{The final publication of this document is available at \url{https://doi.org/10.1016/j.diin.2017.12.004}}

\author[umbcAddress,lpsAddress,bahAddress]{Edward Raff\corref{mycorrespondingauthor}}
\cortext[mycorrespondingauthor]{Corresponding author}
\ead{raff.edward@umbc.edu}
\ead[url]{www.edwardraff.com}

\author[umbcAddress]{Charles Nicholas}
\ead{nicholas@umbc.edu}
\ead[url]{www.csee.umbc.edu/nicholas/charles\_nicholas.html}

\address[umbcAddress]{University of Maryland, Baltimore County}
\address[lpsAddress]{Laboratory for Physical Sciences}
\address[bahAddress]{Booz Allen Hamilton}

\begin{abstract}
Recent work has proposed the Lempel-Ziv Jaccard Distance (LZJD) as a method to measure the similarity between binary byte sequences for malware classification. We propose and test LZJD's effectiveness as a similarity digest hash for digital forensics. To do so we develop a high performance Java implementation with the same command-line arguments as sdhash, making it easy to integrate into existing work-flows. Our testing shows that LZJD is effective for this task, and significantly outperforms sdhash and ssdeep in its ability to match related file fragments and files corrupted with random noise. In addition, LZJD is up to 60x faster than sdhash at comparison time.
\end{abstract}


\end{frontmatter}

\section{Introduction}

In forensic investigations of IT environments, there has been a long recognized and ever increasing need to find similar files for a number of scenarios, including file clustering, detecting blacklisted material, and finding embedded objects \cite{Harichandran2016}. Initial triage and screening of data can easily enter terabytes of data, collected from email archives, hard drives, USB peripherals, and network traffic\cite{Roussev2012}. Such needs occur in many other areas as well, such as firmware analysis \cite{Costin:2014:LAS:2671225.2671232} and malware triage\cite{Jang2011,Lakhotia:2013:VRL:2509225.2509228}. %

Finding similar files is often a daunting task, since manual inspection can take hours per file, if possible at all. The need to automate this task has led to the development of many similarity digests or "hashes" \cite{Breitinger:2013:MNA:2496032.2497148,Breitinger2013a,Winter2013,Kornblum2006,Roussev2011}. Similar to hash functions like MD5 or SHA1, these digests convert an arbitrary string of bytes into a shorter identifying byte string. However, whereas a hash function like MD5 is designed to produce dramatically different output for even a one byte change in the input, these similarity digests are designed to produce little if any change in output given a small change in input. By making the similarity hash insensitive to changes in the input, we can compare the hashes themselves as a method of comparing the similarity of two files.

The two most popular and well known similarity hashes \cite{Harichandran2016} are ssdeep \cite{Kornblum2006} and sdhash \cite{Roussev2010}, which have become the standard benchmarks in the field. While ssdeep is often ineffective for many data types, it is readily available and one of the fastest hashing methods in use. In particular, ssdeep is sensitive to byte ordering, which is a weakness for formats that support arbitrary re-ordering of contents (such as binary executable files). While sdhash is slower than ssdeep, it makes up for runtime performance loss with significantly improved matching and detection ability and is considered state-of-the-art in this regard\cite{Harichandran2016}. The sdhash program's improved matching and detection is the result of resolving the byte reordering weakness of ssdeep. 

\todo[inline,color=yellow]{New paragraph b/c Reviewer two was upset we called ssdeep and sdhash "heuristic". Better? }
While ssdeep and sdhash are popular fuzzy hashing techniques, they have made a number of design or implementation choices that may not be suitable for all the files types we may consider now or in the future. The ssdeep algorithm uses a context triggered approach, and the context itself is dependent both on file length and a minimum block-size $b_min$, and a signature length $S$. Both of these are set to constants without explanation on the determination of these constants, or exploration of their impact. Ssdeep also uses a weighted edit-distance to determine final match scores, without explaining the determination or intuition for the values of the weights\cite{Kornblum2006}. The sdhash algorithm similarly has a number of parameters which must be set, and states they are determined empirically from some set of data\cite{Roussev2010}. However, this data may not accurately reflect the content of interest for practitioners at large, yet the same parameters are now used --- and no tool is provided to re-calibrate such parameters to a desired data type of interest. 
The scoring method used by sdhash
also results in the undesirable property that $sim(A, B) \neq sim(B, A)$ \cite{191669}. It can also be difficult to 
interpret the exact score returned by these methods. For example, \citet{Roussev2012} recommends treating any score in the range of $[21,100]$ as "Strong" in terms of correlation. This covers 77\% of all possible values returned by sdhash. 

In this work, we propose the recently developed Lempel-Ziv Jaccard Distance (LZJD) \cite{raff_lzjd_2017} as an alternative similarity digest. The LZJD was developed for malware analysis, a related field that is particularly challenging due to the arbitrary degrees of freedom available to a malware author. LZJD's success in the area of malware analysis suggests that it may be a useful alternative to sdhash for digital forensic investigations. 

We will show four primary benefits of using LZJD as a similarity digest. First, the time it takes to compare two hashes is orders of magnitude faster with LZJD compared to sdhash, which is critical when dealing with large signature indexes. Second, the LZJD score can in practice be interpreted as a lower bound on how similar the binary contents of two files are. This interoperability is not present in current digest methods. Third, LZJD is better at matching a file fragment with its source file (i.e., the source file receives the highest matching score compared to all other files) compared to both ssdeep and sdhash. 
We suspect that LZJD sets a new state-of-the-art in this regard. Fourth, the digest size of LZJD is fixed, making the determination of index size trivial. 

The rest of this paper is organized as follows. We will introduce the reader to LZJD and its design in \autoref{sec:background_lzjd}. In doing so we will give our interpretation of the LZJD approach that leads us to believe it will make an effective similarity digest. Since efficient execution time is critical to tool adoption and use, we detail how we develop a faster version of LZJD in \autoref{sec:fast_lzjd}, and compare results to the original LZJD work to confirm that our approach has no loss in accuracy while obtaining higher throughput. These tests will also include ssdeep and sdhash to show LZJD's superiority in a related domain, and a significant failure case for sdhash.  Given our new efficient LZJD, we evaluate its abilities as a similarity digest in \autoref{sec:frash_lzjd} using the FRASH framework\cite{Breitinger2013}. We will discuss the meaning and importance of our results in \autoref{sec:discussion}, followed by our conclusion in \autoref{sec:conclusion}.

\section{Background, LZJD} \label{sec:background_lzjd}

The inspiration for LZJD comes from the Normalized Compression Distance (NCD)\cite{Li2004}. The NCD is a metric based on the Kolmogorov complexity function $K(\cdot)$, which returns the length of the shortest possible computer program that could produce a given input string. In measuring the similarity between two byte strings $a$ and $b$, if $a$ has no relationship to $b$ then information about $a$ will not allow us to write a smaller program that produces $b$. Conversely, if $a$ and $b$ are very similar, a program that produces $a$ can also produce $b$ with only a small additional amount of code. The amount of code needed to produce $a$ and $b$ together, rather than independently, is used to measure the similarity of the strings. 

However, the Kolmogorov function $K(\cdot)$ is uncomputable, so an approximation must be used. For the NCD, it was proposed to use any arbitrary compression function $C(\cdot)$, which would return the length in bytes of the compressed input. The resulting distance function is given in \eqref{eq:ncd}, where $xy$ indicates the concatenation of strings $x$ and $y$. 

\begin{equation}\label{eq:ncd}
\text{NCD}(x,y) = \frac{C\left(x y\right)-\min\left(C(x), C(y)\right)}{\max\left(C(x), C(y)\right)}
\end{equation}

Given the widespread availability of compression algorithms, this NCD function is easy to implement in practice. Yet its execution time is hampered by the large amount of time needed to perform compression. The effectiveness of the compression algorithm also has a direct impact on the accuracy of NCD, and the most effective compression schemes often have the greatest run-times. In addition to this computational burden, NCD has a number of practical issues despite theoretical assurances. It has difficulty with high entropy strings, can produce values larger than the theoretical maximum similarity of 1, and lacks symmetry (which also breaks the distance metric properties)\cite{cebrian2005common,Cebrin2007,Borbely2015}. All of these issues would not occur if the Kolmogorov function $K(\cdot)$ could be computed, and occur due to failures in approximating it with a compressor $C(\cdot)$. 

NCD did find use in the domain of malware analysis, where it was found that Lempel–Ziv–Markov chain Algorithm (LZMA) \cite{lzma07url} based algorithms performed best \cite{Alshahwan2015,Borbely2015}. The malware domain often has no obvious or "best" features for general use, and the changing nature of malware over time means that these features can change as well. This makes a method like NCD useful in its flexibility, since no features need to be specified or extracted, and it can work on raw binary contents. However, the lackluster runtime performance of LZMA in NCD limited its use to around 10,000 datapoints or less. 

Inspired by NCD, and noting that the LZMA compression based schemes usually performed best, LZJD was developed to circumvent the performance issues of NCD\cite{raff_lzjd_2017}. LZJD follows a simple process:
First, LZJD converts a byte string $b$ into a set of sub-strings $s_b$, using a simplified version of the Lempel-Ziv 77 algorithm (LZ77) \cite{Ziv1977,raff_lzjd_2017}. This simplified version of LZ77 is presented in \autoref{algo:lz_set}. 

\begin{algorithm}[!htbp]
\caption{Simplified Lempel-Ziv Set\cite{raff_lzjd_2017}}\label{algo:lz_set}
\begin{algorithmic}[1]
\Procedure{LZSet}{Byte string $b$} %
\State $s \gets \emptyset$
\State $start \gets 0$
\State $end \gets 1$
\While{$end \leq |b|$}
	\State $b_s \gets b[start:end]$ 
    \If{$b_s \not\in s$}
		\State $s \gets s \cup \{b_s\} $
        \State $start \gets end$
    \EndIf
    \State $end \gets end +1$
\EndWhile
\State \textbf{return} $s$
\EndProcedure
\end{algorithmic}
\end{algorithm}

We review the details of the LZ77 approach used, as they are important to both our interpretation of score results and in improving the runtime performance of LZJD. This method works by building a set of previously seen strings from the given byte string $b$. The set starts out empty, and a pointer starts at the beginning of the file looking for a sub-strings of length one. If the pointer is looking at a sub-string that has been seen before, we leave it in place and increase the desired sub-string length by one. If the pointer is at a sub-string that has not been seen before, it is added to the set. Then the pointer is moved to the next position after the sub-string, and the desired sub-string length reset to one. Repeating this until no new items can be added to the set, and return the constructed set. The strings in the set will get progressively longer as the length of the input increases.  

Once we have sets of sub-string for each binary of interest, we measure the similarity of the two sets using the Jaccard similarity \eqref{eq:jaccard_sim}, which will return a value in the range of $[0, 1]$. The Jaccard similarity is also referred to as the resemblance between two sets $A$ and $B$. 

\begin{equation}\label{eq:jaccard_sim}
J(A, B) = \frac{|A \cap B |}{|A \cup B|}
\end{equation}

However, this alone is not as fast as desired. So a faster $\text{LZJD}_h$ was introduced\cite{raff_lzjd_2017} to compute approximate similarities. This was done by exploiting the fact that the Jaccard similarity can be computed approximately from a smaller digest produced from the original sets \cite{Broder:1997:RCD:829502.830043,Broder:1998:MIP:276698.276781}. In particular, if $h(\cdot)$ is a hash function, $\text{LZJD}_h$ uses the $k$ smallest hash values as a proxy-set for the original ones, as shown below

\begin{equation}
J(A, B) \approx J\left(\bigcup_{j=1}^k h_{min}^j(A), \bigcup_{j=1}^k h_{min}^j(B) \right)
\end{equation}

Where $h_{min}^n(A)$ indicates the $n$'th smallest hash value from the set $A$. This approximation probabilistically bounds the error at a rate of $O(1/\sqrt{k})$, and the LZJD paper uses $k=1024$ to get an error rate of $\approx3\%$. For the purposes of producing a similarity digest, we note that this hashing scheme makes an excellent candidate for a similarity digest that can be used in the same vein as ssdeep and sdhash. The digest has the benefit of being a fixed maximum size, regardless of the size of the input. We emphasize that the error of this approximation is independent of the size of the input set, and depends only on the size of the digest itself. A direct result is that we can safely choose a fixed digest size that reduces the error down to an acceptable level, and use it for all files \cite{Broder:1997:RCD:829502.830043,Broder:1998:MIP:276698.276781}. We emphasize that this bound allows us to be confident that the estimated Jaccard similarity between two files of disparate size will be close to the correct value. It does not guarantee that LZJD will correctly match disparately sized files.

While the digest will be of a number of elements $k$, the size of the digest on disk may be variable since each item in the LZSet may be a variable number of bytes in length. One might desire a constant digest storage size to make storage planning simpler, and it can also aid in efficient implementations by reducing degrees of freedom (which will allow for more performance optimization). We achieve this in this work with our design of a faster implementation of LZJD, which we will detail in \autoref{sec:fast_lzjd}, and show that we are able to obtain a digest with fixed storage size and considerable performance improvements without compromising on the accuracy of LZJD. 

We argue that the grounding in Jaccard similarity approximations is also more interpretable than the scores produced by ssdeep and sdhash. 
For a direct interpretation of the math behind the LZJD score, consider two inputs $A$ and $B$. A score of 0.75 means that, for all sub-strings shared between the LZSet($A$) and LZSet($B$), 75\% of them could be found in both files. This can be loosely interpreted as saying that $A$ and $B$ share 75\% of their byte strings. 
This is not an exact measure of byte content similarity, and will be impacted by two primary factors. First, that the hashing of sub-strings does not attempt to maintain information about string length. We expect this to be approximate to the average string length over many hashes, but this will introduce variability in the scores. Second, that the LZ set creation can be impacted by the contents of the binary, so it is possible to produce different sets for similar inputs. We will see that this issue does impact the score returned, but does not seem to reduce the matching ability of LZJD. We also note that sdhash has a similar issue where inputs can be modified by an adversary to reduce the matching score\cite{Breitinger2012}, but has found widespread use regardless. So we do not believe this potential shortcoming would be a hindrance in practice. 

An approximate (empirically observed) bound can be given to this interpretation by noting a unique property of the FRASH framework. 
For each test, we can analytically determine what the Levenshtein distance \cite{levenshtein1966binary}, or edit-distance, between files would have been in each test. The edit-distance being the minimum number of operations needed to transform one string into another, where an edit can either replace, remove, or add a byte to the string. The edit distance between two binary files would not normally be %
computationally feasible, as it is an $O(n^2)$ cost to determine this value for two strings of length $n$. Because the FRASH tests alter the binaries in a specific way for each test, we know the edit-distance between the original file and the modified versions created by FRASH. 

With this insight, we find that LZJD tends to act as a lower bound of \eqref{eq:lzjd_lb}

\begin{equation}\label{eq:lzjd_lb}
J(\text{LZSet}(A), \text{LZSet}(B)) \lesssim \frac{\text{edit-distance}(A,B)}{\max\left(|A|, |B|\right)}
\end{equation}

We use the approximately less-than symbol $\lesssim$ because this is not a proven bound, and does not hold for every experiment. Equation \eqref{eq:lzjd_lb} is an empirical observation that we discuss further in \autoref{sec:discussion}. While it is satisfied by the majority of tests in FRASH, it does not hold for all of them. We argue that this interpretation may be useful for practitioners. 

Ultimately, the $\text{LZJD}_h$ similarity/distance performed orders of magnitude faster than NCD, with equal or better accuracy, on several malware datasets for both malware detection (correctly labeling a binary as benign or malicious) and malware family detection (finding the correct malware family for a known malicious binary). This success, combined with its use of a fixed-length digest for faster distance computations, inspires our hypothesis that it could be successfully used for the same kind of digital forensic scenarios as ssdeep and sdhash. We evaluate this feasibility in \autoref{sec:frash_lzjd}. But first, we must further improve the runtime efficiency of LZJD to make it practical for this application. 

\section{A Faster LZJD Implementation} \label{sec:fast_lzjd}

We now review the high level details of the original LZJD implementation, and discuss our modifications that result in a faster variant appropriate for the forensic use case. This implementation is in Java, and we note that both ssdeep and sdhash are written in C/C++. This may mean that there is still room for improved performance of our new LZJD implementation. We have made a Java implementation
\footnote{\url{https://github.com/EdwardRaff/jLZJD}} 
of this faster LZJD available to the public. The program has the same command line arguments as sdhash in order to facilitate integration with existing work flows. 
We are also working on a C++ version
\footnote{\url{https://github.com/EdwardRaff/LZJD}} 
, though performance optimization is not yet complete

The original version of LZJD was a rather naive Java implementation. The set $s$ in \autoref{algo:lz_set} was a simple HashSet of ByteBuffers. A ByteBuffer object represents a byte string. This choice meant that equality comparisons had to compare each byte in each buffer, which would take time linear with respect to the current sub-string under consideration. Furthermore, and to the detriment of performance, these comparisons force the hash of the string to be re-computed at every step, resulting in redundant work.

Once the set of ByteBuffers was obtained, the MD5 hash of each member in the set was computed and the lower 32 bits used for the min-hashing. This set of integers was then sorted, and the minimum $k$ integers created the final set used for this faster variant of LZJD, which we will denote as \lzjdh. The MD5 function was chosen to ensure even distribution of hash values, which are the result of its original design as a cryptographic hash function. 

We will present tests in \autoref{sec:lzjd_speedup_results} that show these modifications do not degrade the accuracy of LZJD but do significantly reduce the runtime cost. 
We do this by performing hashing continuously as data is read in, and representing every sub-string by the hashed integer counterpart. By using a hash that we update with one byte at a time, we no longer need to read the entire file into memory for \lzjdf to work. This may result in false collisions during the LZ set construction as two hashes may collide to the same integer, but we believe the cost of such collisions to be minimal. The LZ algorithm will simply continue processing the next byte, which is now a new sub-string that is one byte longer. It is necessary that this new sub-string does not currently exist in the set, because the previous set did not contain the true prerequisite sub-string either. For the new sub-string to also have a collision becomes astronomically unlikely, assuming the hashes are uniformly distributed. Even if several collisions occurred, the impact on the output similarity should be minimal, as the sub-strings of each sub-string are also in the sets and included in the comparison. That is to say, if the sub-string "abcdefg" is not included in the set due to a hash collision, the contributions of "abcdef", "abcde", etc., are still present. 

To make sure these hash values are of a high quality, but avoiding the unnecessary quality of a  hash function like MD5, we use the MurmurHash3\footnote{\url{https://github.com/aappleby/smhasher}} function. This hash function is designed to have an even distribution of hashes and require minimal CPU time for computation. While not originally designed for it, we re-implement this algorithm so that the hash can be updated one byte at a time. This requires keeping a four-byte memory that is updated and used to compute the running hash output, in addition to the internal state of the MurmurHash3 algorithm. 

We also optimize the integer set object to take advantage of the two unique artifacts of the situation. First, it only needs to support the insertion of integers, so no removals are needed. Second, since the integer values are hashes, there is no need to apply any kind of hash function to them, as they will already be evenly distributed (i.e., our hash set can use the identity function as its "hash" function). We thus adapt an open addressing scheme with double hashing \cite[p. 528--529]{Knuth:1998:ACP:280635} that is normally used for a hash table. We can reduce memory use by ignoring the "value" part, and using a boolean array to indicate if an entry is free or filled, and remove logic normally needed to handle the removal of entries. The "key" alone will then act as the set entry, with an implicit null "value". This reduces memory use and execution time.

Once the entire file is processed, we will have a set of integers, which we will then convert to a list of integers. Rather than naively sorting the list, which is $O(n \log n)$, we instead apply one of many algorithms that returns us the $k$ smallest items in $O(n)$ time \cite{Dor:2001:MSR:587900.587932}. Beyond optimizing how the set of $k$ values is obtained, we can further improve how they are stored and compared. 

The original LZJD would store the set of $k$ integers in a set object, and to compute the size of the intersection of two sets, would iterate over one set and query for its entries in the other. This results in $O^*(k)$ time complexity, but is both memory inefficient and results in random memory access that negatively impact cache and pre-fetching performance. Instead we store the $k$ items in a sorted array, which is $O( k \log k)$, but $k << n$, so this sort is of minor impact. The benefit is that we can compute the intersection by doing a merge-sort like comparison of the values in each array, incrementally stepping forward in one list when its value is less than another. This well-known approach is given in \autoref{algo:sorted_intersection}, and results in a non-amortized  $O(k)$ runtime for digest comparisons. Further, the dense arrays are more memory efficient, and the incremental walk through the sorted arrays will work \textit{with} the hardware pre-fetching for improved performance.  

\begin{algorithm}[!htbp]
\caption{Set Intersection Size via Sorted Lists}\label{algo:sorted_intersection}
\begin{algorithmic}[1]
\Procedure{Intersection}{Integer arrays $a$ and $b$} %
\State $\text{pos}_a \gets 0, \text{pos}_b \gets 0$
\State $size \gets 0$
\While{$\text{pos}_a < |a|$ and $\text{pos}_b < |b|$ }
	\If{$a[\text{pos}_a] < b[\text{pos}_b]$}
    	\State $\text{pos}_a \gets \text{pos}_a + 1$
    \ElsIf{$a[\text{pos}_a] > b[\text{pos}_b]$}
    	\State $\text{pos}_b \gets \text{pos}_b + 1$
    \Else \Comment{Equal values, means item was in both}
    	\State $\text{pos}_b \gets \text{pos}_b + 1$
        \State $\text{pos}_a \gets \text{pos}_a + 1$
        \State $size \gets size +1$
    \EndIf
\EndWhile
\State \textbf{return} $size$
\EndProcedure
\end{algorithmic}
\end{algorithm}

\subsection{LZJD Speedup Results} \label{sec:lzjd_speedup_results}

Having specified the modifications that produce the faster \lzjdf, it is important to validate that the hashing approach does not meaningfully degrade accuracy compared to the original \lzjdh. To do so, we will repeat the malware family classification experiments used in \cite{raff_lzjd_2017}. The malware classification problem has been previously identified as an area whether similarity digests could be useful\cite{Harichandran2016}, making this test of particular relevance in this context of similarity digest comparisons. For this reason we will also include ssdeep and sdhash in this comparison, and see that \lzjdf outperforms them both. 

Malware family classification can be seen as a close corollary to the digital forensics problem of finding a related file. For each malware sample, we wish to identify the family it belongs to by comparing the sample to a database of known malware. Each specimen in the same malware family is intrinsically similar, and can be seen as one unit of "sameness" for which the inter-family similarity should be higher than the similarity to any other arbitrary sample. This task is strongly correlated with matching a modified file to its original file, but can be seen as a more challenging scenario. This is because malware is often written by an active adversary which attempts to avoid detection. Metamorphic malware, which changes itself upon propagation, makes this a common and difficult scenario \cite{Wong2006,RHUL-MA-2008-02}. 

The two malware datasets used each have two variants of the experiment. The Microsoft malware comes from a 2015 Kaggle competition, and the data is provided and labeled by Microsoft \cite{microsoft_kaggle_2015}. There are 9 malware families in 10,868 files. The first variant of this dataset uses only the raw bytes of the original files, with the PE-header removed\footnote{The PE header info was removed by Microsoft to avoid accidental infection, and cannot be reversed.}. The raw binaries take 50.8 GB of storage space, and we will refer to this dataset as "Kaggle Bytes". The second variant is the disassembly produced by IDA-Pro, which is a more human-readable version of the files. This variant takes up 147GB of space, and we will refer to this dataset as "Kaggle ASM". 

The second dataset is Android malware from the Drebin corpus \cite{Arp2014}. Following \cite{raff_lzjd_2017}, we remove any malware family that had less than 40 samples. This results in a dataset with 20 malware families and 4664 samples. Android applications are normally distirbuted as APKs, which are simply zip-files. Because the compression applied by zipping the contents can impact the effectiveness of our hashes, we evaluate the dataset in two ways. One using the raw APKs ("Drebin APK"), and the other using an uncompressed tar of the APK contents ("Drebin TAR"). These variants take 6.4GB and 8.6GB respectively. Differences in performance between these two datasets can be wholly attributed to the impact of compression\footnote{We note that the amount of compression applied to the APKs is generally light, as  a trade-off is being made between storage size and power consumption, both limited resources on mobile phones}, since it is the only source of variation between the two sets. 

We also note the importance of these tests in regards to the performance of LZJD and other tools in high-entropy situations. LZJD was analytically predicted to experience sub-optimal behavior when encountering high entropy data, yet empirically performed well when given such data \cite{raff_lzjd_2017}. The impact of high entropy is discussed further in the FRASH test in \autoref{sec:scb}, which use random bytes as part of the test to increase the matching challenge. The Kaggle and Drebin datasets help to validate that LZJD works even when high entropy is present, with the Android APK corpus having a median byte entropy of 7.96. Thus the performance of LZJD, ssdeep, and sdhash in this tast can be seen as a test of all three approaches when dealing with higher entropy content. 

To evaluate all of our hashing options on this dataset, we will use 10-fold cross validation. We will use the 1-nearest neighbor algorithm to classify each sample against the other folds. If the matching algorithm returns the highest similarity score for a member of the same malware family, then the algorithm correctly classified that point.  For each fold we will measure the balanced accuracy \cite{Brodersen:2010:BAP:1904935.1905533}. The balanced accuracy gives equal total weight to each class. This is useful since the malware families are not evenly distributed, and results would be skewed upward by the most populous families. The accuracy for each method on each dataset is presented in \autoref{tbl:mfc_results}.

\begin{table}[!htb]
\centering
\caption{Balanced accuracy results on each data and feature set. Evaluated with 10-fold CV, standard deviation in parenthesis. }
\label{tbl:mfc_results}
\begin{adjustbox}{width=\columnwidth}
\begin{tabular}{@{}lcccc@{}}
\toprule
Dataset         & ssdeep (\%) & sdhash (\%) & \lzjdf (\%) & \lzjdh (\%)         \\
\midrule
Kaggle Bytes    & 38.4 (1.4)             & 60.2 (2.3)              & 98.0 (1.2)              & 97.6 (1.5)                       \\
Kaggle ASM      & 26.6 (2.2)             & 28.8 (1.3)              & 96.7 (1.9)              & 97.1 (2.0)                       \\
Drebin APK      & 13.6 (1.6)             & 5.8 (0.5)               & 81.3 (4.6)              & 80.8 (2.6)                       \\
Drebin TAR      & 24.2 (2.9)             & 8.3 (1.2)               & 87.5 (2.0)              & 87.2 (2.8)                       \\
\bottomrule
\end{tabular}
\end{adjustbox}
\end{table}

Here it is easy to see that our new \lzjdf does not meaningfully change the performance on these datasets compared to the original \lzjdh. The largest change is an increase in standard deviation on the most difficult dataset (Drebin APK). However \lzjdf has slightly higher mean accuracy and lower standard deviation on most of the datasets. This closeness in results indicates the high fidelity of our new approach, and that the simplifications in LZSet implementation do not meaningfully impact the quality of results. This gives us confidence that our changes to \lzjdf will generally perform well.

Comparing both LZJD implementations to ssdeep and sdhash, we can see far superior classification accuracy. The closest either ssdeep or sdhash come to matching LZJD's performance is on the Kaggle Bytes dataset, where sdhash still trails by over 37 whole percentage points. We will see this trend of LZJD having superior matching ability repeated in \autoref{sec:frash_lzjd}. 

While sdhash performs better than ssdeep on the Kaggle datasets, we also see sdhash produce degraded results on the Drebin datasets. Its scores of 5.8\% and 8.3\% accuracy are barely better than the 5\% threshold for random guessing. When inspecting these results manually, we discovered that the root cause is related to the %
nature of sdhash's scoring algorithm. Sdhash ends up keying off features generally common to all of the Android samples in our corpus, producing average nearest neighbor scores of 99.7 and 99.9 for Drebin APK and Drebin TAR respectively. This use case provides credence to the desire for a more principled and interpretable score function.

\begin{table}[ht]
\centering
\caption{Statistics on the direct, absolute, and relative differences between \lzjdh and \lzjdf similarities for all pairwise distances. Scores for the first four columns are out of a maximum score of 100 for the difference. The last two columns are shown in percentage points. }
\label{tbl:lzjd_hf_diffs}
\begin{adjustbox}{width=\columnwidth}
\begin{tabular}{@{}lcccccc@{}}
\toprule
             & \multicolumn{2}{c}{Difference} & \multicolumn{2}{c}{Absolute Difference} & \multicolumn{2}{c}{Relative Difference} \\ 
             \cmidrule(l){2-3} \cmidrule(l){4-5}  \cmidrule(l){6-7} 
Dataset      & Avg.         & Stnd. Dev.      & Avg.             & Stnd. Dev.           & Avg. (\%)        & Stnd. Dev. (\%)          \\ \midrule
Kaggle Bytes & 0.231        & 0.871           & 0.647            & 0.627                & 0.755            & 0.864                \\
Kaggle ASM   & 0.010        & 0.793           & 0.531            & 0.588                & 0.601            & 0.783                \\
Drebin APK   & 0.010        & 0.691           & 0.489            & 0.489                & 0.539            & 0.660                \\
Drebin TAR   & -0.056       & 0.623           & 0.450            & 0.434                & 0.491            & 0.624                \\
t5           & 0.112        & 0.505           & 0.332            & 0.397                & 0.351            & 0.445                \\ \bottomrule
\end{tabular}
\end{adjustbox}
\end{table}

To further confirm the high fidelity of \lzjdf's approximation of \lzjdh, we also look at the statistics of all pair-wise distance computations in each dataset, and include the t5 corpus that will be further discussed and used in the next section of this work. We will look at three sets of statistics, where $d_h = \text{LZJD}_h(A,B)$ and $d_f = \text{LZJD}_f(A,B)$. First at the average difference, $d_h-d_f$, which we want to see centered around zero (indicating the approximation is unbiased in practice). The average absolute difference, $|d_h-d_f|$, which we wish to see being as small as possible (indicating the approximation is accurate). Last, we will consider the relative difference, $|d_h-d_f|/\max\left(d_h, d_f, 0.01\right)$, which helps us further consider changes based on their relative magnitudes. We add the $0.01$ term to the relative difference computation to avoid division by zero, which occurred when there was no error in the approximation of files with no similarity. 

The average and standard deviations for these three statistics are presented in \autoref{tbl:lzjd_hf_diffs}, where the maximum possible difference would be 100. We can see that the average relative difference is less than a percentage point, with and at three standard deviations out is still less than a 4\% error. Similarly the worst average absolute difference indicates that the majority of scores will differ by no more than 4 points out of 100. We also see that the average total difference is centered around zero. These results give us clear validation that our \lzjdf approximation is not only faithful to the true nearest-neighbor ordering provided by \lzjdh, but also accurately reproduces the same score values. That is to say, we have empirically observed that $|d_f - d_h| < \epsilon$. 

\begin{table}[ht]
\centering
\caption{Total evaluation time for each method in performing 10-fold CV. Time presented in seconds. }
\label{tbl:mfc_time_results}
\begin{adjustbox}{width=\columnwidth}
\begin{tabular}{@{}lcccc@{}}
\toprule
Dataset         & ssdeep                & sdhash                 & \lzjdf                  &  \lzjdh         \\
\midrule
Kaggle Bytes    &  \num{3.02e3}         &   \num{8.64e5}         &   \num{3.17e3}          & \num{1.73e4}                    \\
Kaggle ASM      &  \num{3.25e3}         &   \num{4.74e6}         &   \num{1.44e4}          & \num{4.85e4}                    \\
Drebin APK      &  \num{2.21e2}         &   \num{1.30e4}         &   \num{5.56e2}          & \num{7.17e3}                    \\
Drebin TAR      &  \num{2.76e2}         &   \num{2.04e4}         &   \num{6.46e2}          & \num{7.65e3}                    \\
\bottomrule
\end{tabular}
\end{adjustbox}
\end{table}

\begin{table}[ht]
\centering
\caption{Time spent hashing for each method in performing 10-fold CV. Time presented in seconds. }
\label{tbl:mfc_hashing_time_results}
\begin{adjustbox}{width=\columnwidth}
\begin{tabular}{@{}lcccc@{}}
\toprule
Dataset         & ssdeep                & sdhash                 & \lzjdf                  &  \lzjdh         \\
\midrule
Kaggle Bytes    &  \num{5.42e2}         &   \num{1.72e3}         &   \num{1.92e3}          & \num{1.22e4}                    \\
Kaggle ASM      &  \num{2.26e3}         &   \num{5.34e3}         &   \num{7.44e3}          & \num{4.11e4}                    \\
Drebin APK      &  \num{1.85e1}         &   \num{2.35e2}         &   \num{3.74e2}          & \num{4.99e3}                    \\
Drebin TAR      &  \num{2.51e1}         &   \num{3.52e2}         &   \num{4.36e2}          & \num{5.47e3}                    \\
\bottomrule
\end{tabular}
\end{adjustbox}
\end{table}

To evaluate the runtime of our new \lzjdf, we can see the total time taken for both hashing the files and performing the nearest neighbor searches in \autoref{tbl:mfc_time_results}. As desired, we can see that \lzjdf is consistently faster than \lzjdh, by a factor of 3.4 to 12.9. We can further see that this total evaluation time is comparable to ssdeep, and generally two orders of magnitude faster than sdhash. These large speed advantages generally come from \lzjdf being faster to compare. These results %
support our claim that our new \lzjdf is fast enough to be a practical alternative to both ssdeep and sdhash. 

In \autoref{tbl:mfc_hashing_time_results}, we show the time spent creating the digests for this same task. This allows us to see that for ssdeep, sdhash, and \lzjdf, creating the digest itself is usually small relatively to the amount of time spent. In these tests data was read from hard disk, and we see that \lzjdf takes 12\% to 60\% more time to create the digest compared to sdhash. Given that the total time for \lzjdf is orders of magnitude faster than sdhash, this allows us to confirm that the fast comparison time is the source of this dramatic speed advantage. We will explore the performance differences further in \autoref{sec:frash_efficiency}. 

Looking at the hash time also demonstrates the critical importance of our optimization toward practical use. For \lzjdh, the hashing time is one to two orders of magnitude greater than our improved \lzjdf.  \lzjdh is the only metric which spends the majority of time in producing the digests itself for every dataset. These optimizations were thus necessary to make the tool usable for practitioners, with digest time comparable to sdhash while providing faster digest comparison. 
For the remainder of the paper, we will simply refer to \lzjdf as LZJD for brevity.

\section{Similarity Hash Comparisons using FRASH} \label{sec:frash_lzjd}

The evaluation of similarity digests is not a trivial matter.
It requires a diversity of file types (that should reflect real world content) and some level of ground-truth about which files are similar to others. 
\citet{Roussev2011} introduced the \textbf{t5} corpus for such evaluations\footnote{available at \url{http://roussev.net/t5/t5.html}}, and a manual evaluation of sdhash was performed. The t5 corpus contains a number of different file types, summarized in \autoref{tbl:t5_file_types}. \citeauthor{Roussev2011} also proposed a number of challenges for which one would want to use a similarity hash, which \citeauthor{Breitinger2013} drew from to create the automated FRASH test suite \cite{Breitinger2013}. FRASH is not exhaustive of all the ways in which one may use a similarity digest, but it does provide a valuable reference point across a spectrum of common use cases. Similarly, the widely used t5 corpus is only a subset of potential file types and data sizes that one may encounter. 

\begin{table}[!htbp]
\centering
\caption{Contents of the t5 corpus. There are 4475 files in total, totaling 1.9 GB in size.}
\label{tbl:t5_file_types}
\resizebox{\columnwidth}{!}{\begin{tabular}{@{}lcccccccc@{}}
\toprule
                & html & text & pdf  & doc & ppt & xls & jpg & gif \\ \midrule
Number of Files & 1093 & 711  & 1073 & 533 & 368 & 250 & 362 & 67  \\ 
Avg. File Size (KB)  & 66 & 345  & 590 & 433 & 1003 & 1164 & 156 & 218  \\ 
\bottomrule
\end{tabular}}
\end{table}

In particular, the FRASH tests  evaluate four desirable qualities:
\begin{enumerate}
\item \textit{Document similarity detection}: where we wish to determine which documents are intrinsically related, such as multiple revisions of the same word document. 
\item \textit{Embedded object detection}: where the goal is to detect that one object type (such as an image) has been placed inside of another (such as a email document). 
\item \textit{Fragment detection}: where we are given a sub-string of some larger file, and we wish to identify the source of this sub-string. 
\item \textit{Clustering files}: where we wish to group similar files together. 
\end{enumerate}

The FRASH suite is written in Ruby\footnote{FRASH is available at \url{http://www.fbreitinger.de/wp-content/uploads/2017/04/FRASH_1.01.zip}}, and allows for easy integration of new similarity hashing schemes. The tests are divided into two higher level sections. The first section is \textit{Efficiency}, which measures only runtime properties of the hash digest. This includes the digest time, comparison time, and hash size relative to the input. Our improvements to the LZJD algorithm tackle only these quantities, which are critical when up to terabytes of data may need triaging\cite{2009:BBS:1488734.1490103,Roussev2012}. 

The second, and more expansive, are the \textit{Sensitivity \& Robustness} tests. These evaluate the ability of the hash function to perform matching under various circumstances, and the quality of the match score returned in each scenario. These tests will show that LZJD possesses a superior ability to correctly match a fragment to the correct source file, even when presented with significant byte alterations or comparatively small fragment sizes. 

Below we will present and discuss the results from each of the tests in the FRASH suite. For each result we will only present a portion of the output for brevity and readability, with the algorithm getting the most successful matches shown in \textbf{bold} for each test. More complete results can be found in \ref{sec:full_frash}. All tests were run on a computer running OSX version 10.10.5. With a 2.66 GHz Intel Core i5 CPU and 16 GB of RAM. In initial testing, the FRASH code was highly sensitive to random read/write time. Initial runs on a standard HDD resulted in runtime that would take days for as few as 20 files. For this reason, all code and data used were stored in a RAM Disk. This is a method by which a virtual disk is created on the system that acts likes any other file system, but all stored files are kept only in RAM. This avoided all issues with the random access impact on test runtime. 

\subsection{Efficiency} \label{sec:frash_efficiency}

In this section we are concerned with the computational and storage efficiency of each hashing method. This is measured by computing the hash digest for every file in the t5 corpus, and creating a digest file containing every hash. Once complete, all $n^2/2$ pairwise distance computations are done. This allows us to measure the runtime efficiency of the hashing process as well as the comparison of hashes, and the storage efficiency of the hash size itself. For only the efficiency tests, the SHA1 hash function is included by the FRASH suite as a benchmark for both time and space. The intuition for comparing with the SHA1 hash is that it serves as a useful barometer for grounding the compute efficiency of storage cost of these digests for those less familiar with them. 

\begin{table}[!htbp]
\centering
\caption{Runtime efficiency results. Time taken to compute all hashes for each method, and the time needed to perform all-pairs distance computations. All times measured in seconds using only one CPU core. Best results shown in \textbf{bold}, and second best in \textit{italics} (ignoring SHA baseline). }
\label{tbl:hashing_time}
\begin{adjustbox}{max width=\linewidth}
\begin{tabular}{@{}lcrrc@{}}
\toprule
        & Average & \multicolumn{1}{c}{Total} & \multicolumn{1}{c}{All-pairs} & $\text{SHA1}^{-1}$     \\ \midrule
sha1sum & 0.01519 &  67.7                     & ---                           & 1.00                    \\
ssdeep  & 0.01223 & \textbf{54.5}                      & \textit{32.1}                          & 0.81                    \\
sdhash  & 0.04241 & 189.0                     & 496.5                         & 2.79                    \\
LZJD    & 0.03159 & \textit{140.8}                     & \textbf{8.2}                           & 2.08                    \\ \bottomrule
\end{tabular}
\end{adjustbox}
\end{table}

The runtime results can be found in \autoref{tbl:hashing_time}, where we see the average and total time taken to hash the files of the t5 corpus. The total hashing time (second column) is measured using the Unix time command when giving the t5 corpus as the only input for each hashing implementation
. 
The rightmost column shows how many times longer each method took to compute all hashes compared to the SHA1 hash. Here we can see that sdhash is the slowest hash function by a factor of 2.8, and that our Java LZJD implementation is 34\% faster than sdhash. Ssdeep is the fastest at 23\% faster than than the SHA1 algorithm, but lacks in its ability to perform accurate matching once the hashes are produced.

We note that while the FRASH test showed LZJD was faster in hashing time compared to sdhash, our test sin section \autoref{sec:lzjd_speedup_results} showed it to be slower. The difference between these tests is the use of the RAM disk for FRASH to avoid prohibitive test times due to random accesses. These results combined would indicate that sdhash and LZJD are roughly comparable in hashing time, and we may expect to see variation in which one is faster based on unique hardware combinations and situations. 

LZJD's runtime performance is better still when we look at the time needed for comparing the hash outputs, and is over 60 times faster than sdhash in this respect, and still 3.9 times faster than ssdeep. This would indicate that LZJD would be preferable in a situation where we have many known objects of interest in a database, and need to process the contents of a new device against the known database. 
One might argue that 
having a faster digest comparison is more important than a faster digest calculation.
Indeed, others have worked on building special indexes specifically to accelerate the bottleneck of comparing many digests\cite{Winter2013}. 

The time needed to hash $n$ files is naturally an $O(n)$ task, but comparing the $n$ derived hashes to an existing database of $m$ hashes is $O(n m)$ in complexity. The latter will clearly become the dominant cost as the number of objects under consideration increases, and so we would want to minimize its base time requirements as much as possible. An example of this can be found in \cite{Costin:2014:LAS:2671225.2671232}, where sdhash and ssdeep were used on 1.2 million files extracted from firmware images. Due to the computational burden at comparison time, these hashes couldn't be applied to the entire corpus. Our results in \autoref{sec:lzjd_speedup_results} corroborate this high comparison time cost, where we see LZJD compare favorably to both ssdeep and sdhash. LZJD's efficient digest comparison pushes back this limitation. 

We observe that this issue of runtime efficiency has been noted before, and others have attempted to build more efficient indices for specific use cases. \citet{Winter2013a} built an indexing scheme for the ssdeep algorithm, but ssdeep's low precision and recall limit the utility of such a tool. \citet{Breitinger2014} build a more general purpose index that is compatible with sdhash, but cannot return or filter based on similarity scores or indicate which specific file as a match. 
This work was later extended to resolve these issues, allowing it to return exact file matches \cite{DBLP:journals/jdfsl/BreitingerRB14,Lillis2017}. While able to obtain speedups of up to a factor of 2.6, it does not guarantee all matches will be found. 

LZJD provides a sound method of circumventing these issues that may be explored in future work. Since LZJD is a valid distance metric, it avails itself to more principled and existing indexing strategies that are designed for metric spaces. These indices support $O(\log n)$ query time\cite{Yianilos1993,Uhlmann1991,Beygelzimer2006,Izbicki2015} and guarantee that all neighbors will be found. 

\subsubsection*{Parallel Computation}

All of the runtime performance numbers addressed so far, and that we will discuss for the remainder of this work, are with respect to single-core execution. We take a brief moment to mention that both sdhash and our LZJD implementation support parallel computation. Effective speedup will be a factor of the number of cores available, as well as the amount of data being processed. The more data being computed on, the easier it is to effectively use more cores. 

Some parallel performance tests were run on a large multi-core server to test both sdhash and LZJD's scalability. This server has four Intel Xeons with 20 cores each, for a total of 80 CPU cores, 2 TB of RAM, and 40 TB of SSD storage. Single threaded on this machine, sdhash took 77 seconds to create a digest of the t5 corpus, and 262 seconds to perform the all-pairs comparison. LZJD took 79 seconds for the digest, and and 6.7 seconds for the all-pairs comparison. 

Overall, testing on this machine indicates that sdhash and LZJD have approximately the same scaling as more cores are used. Using all 80 cores, sdhash had a digest speedup of 13.0x and LZJD 13.6x. While the digest problem is trivially parallelizable, reading the data from disk becomes the bottle neck and limits scalability. For the digest all-pairs comparison, sdhash received a speedup of 6.5x and LZJD of 5.1x. In this case scalability is likely limited by memory bottlenecks, as it is difficult to keep 80 cores adequately primed with a relatively small amount of data. In the case of LZJD, at 5.1x speedup the time for LZJD all-pairs is just 1.3 seconds, most of which is spent reading in the digest file from disk. We also note that LZJD's single threaded time  of 6.7 seconds is already faster than the 40.2 seconds sdhash takes when using all 80 cores.

\subsubsection{Compression}

The efficiency test in FRASH also produces a set of compression results. These results are concerned with the size of the hash digest with respect to the original file sizes. All things being equal, it is preferable to have a digest that is smaller rather than larger. A smaller digest size allows for the storage and transport of larger databases, and gives some indication about the information efficiency of the digest itself. 

\begin{table}[!htbp]
\centering
\caption{Compression test results. First column shows the average length of the digest, followed by the average ratio between digest length and original file length. The last two columns show the maximum ratio encountered and the size of the entire digest for all files. }
\label{tbl:compression}
\begin{adjustbox}{max width=\linewidth}
\begin{tabular}{@{}lcccc@{}}
\toprule
        & Avg. Length & Avg. Ratio (\%) & Max Length (ratio)  & Digest Size \\ \midrule
sha1sum & 20 B        & 0.0047          &   ---               & 420 KB           \\
ssdeep  & 57 B        & 0.0133          & 78 B ($\leq$0.01\%) & 592 KB           \\
sdhash  & 10.6 KB     & 2.5203          & 409 KB (2.93\%)     & 61.3 MB          \\
LZJD    & 4.01 KB     & 0.9566          & 4.01 KB (10.1\%)    & 23.5 MB          \\ \bottomrule
\end{tabular}
\end{adjustbox}
\end{table}

The compression results are shown in \autoref{tbl:compression}, where the first two columns present the average length of the digest, and the average percentage of the digest size with respect to the original file. Here we can see that SHA1 and ssdeep both produce very small digests. Sdhash produces the largest digests, with an average of 10.6 KB that is usually 2.5\% of the original file size. LZJD falls in a middle ground, with an average digest of 4 KB, 2.65 times smaller than sdhash. By the nature of our LZJD hash, the digest size will never be more than 4 KB\footnote{With minor overhead for the header matching sdhash's output style.}. Smaller digests may occur for small files which can be represented with less than 1000 dictionary entries for the Lempel-Ziv process.  This makes LZJD especially effective for large files, with theoretical support for its method of production. However, LZJD's fixed size can also result in an overly large digest for small files, as can be seen by the maximum digest-size to original-size ratio of 10\%. 

\subsection{Sensitivity \& Robustness}

We will now review the Sensitivity \& Robustness tests that are a part of the FRASH framework. Tests \autoref{sec:scb} and \autoref{sec:rand_resist} will run a digest comparison on only two files at a time, namely a source file and a target file. The source file will be an unaltered file from the t5 corpus. The target file will be a modified version of the source file. These tests will be measuring behavior of the scoring methods used and how they change with changes to a single file. The implicit assumption of the FRASH framework is that a higher score between two matching files is \textit{always} better, all other things being equal. As we discussed in \autoref{sec:background_lzjd}, the LZJD score will be based on the amount of byte similarity -- and will not attempt to reflect "match or no match" as ssdeep and sdhash do. This makes comparing the results in these tests more challenging, and we will discuss the issue of score function behavior further in \autoref{sec:discussion}. 

Tests \autoref{sec:frag} and \autoref{sec:align} will generate a digest database from the whole t5 corpus, and then see if a target file (still a modified version of one of the t5 source files) can be correctly matched to its source. In these tests the goal is for us to correctly match a file to its source, and can be viewed as equivalent to the nearest neighbor problem we visited in \autoref{sec:fast_lzjd}. These tests can be thought of as a harder variant of the task from a machine learning sense, as there is only one correct neighbor for each test point (which would be the source point), where any file from the same class would be considered correct for the tests done in \autoref{sec:fast_lzjd}. We will see that LZJD presents a new state-of-the-art in matching ability, far  exceeding both ssdeep and sdhash in its matching ability. 

\subsubsection{Single Common Block correlation} \label{sec:scb}

The Single Common Block (SCB) test is designed to determine how small a "common block" of identical content can be before a digest algorithm produces a score of zero (i.e., no commonality). This test compares only two files at a time, where each file has random byte contents. A portion of each file will be set to the same common content, and this common block will be iteratively decreased in size. This test was run 50 times with 50 different source files to extract common blocks from. In the original FRASH testing, it was found that sdhash was able to produce matches for smaller common blocks then ssdeep, but ssdeep was able to produce higher matching scores. 

For the tables in this section, the \textit{Average Block size (KB)} indicates how small the common object's size was to reach a score greater than or equal to a minimum score threshold. Similarly, \textit{Average Block size (\%)} is how small this single common object was as a percentage of the block size. The \textit{Matches} line in each table is the number of files (out of the 50 selected) that were able to be matched and achieve a score at or above the given score. The two aforementioned averages are with respect to the files matched at that level. 

This particular test puts LZJD at a disadvantage, because its score does not have the same meaning as sdhash and ssdeep, and because the files are produced with completely random byte sequences. Random bytes are a weakness of LZJD in the case of matching similarity, because the LZ algorithm will begin collecting all smallest sub-strings, which will cause a non-zero match to occur. This makes it impossible to reach the original termination case of FRASH, and we we terminate LZJD in this test after a SCB size of 16KB. Thus, when interpreting \autoref{tbl:scb_2mb} and \autoref{tbl:scb_512kb}, the score that has an average block size of 16 KB should be treated as the same as the zero score for sdhash and ssdeep. 

\begin{table}[!h]
\centering
\caption{Single Common Block results for a 2MB file. Columns show the scores achieved, and rows the size of the common block of content, and the number of successful matches (max of 50).  }
\label{tbl:scb_2mb}
\begin{adjustbox}{max width=\linewidth}
\begin{tabular}{@{}clccccc@{}}
\toprule
                        & Score                 & $\geq$25 & $\geq$15  & $\geq$10  & $\geq$5 & 0     \\ \midrule
\multirow{3}{*}{\rotatebox[origin=c]{90}{\footnotesize ssdeep}}
                        & Avg. block size (KB)  & 386   & ---   & ---   & ---  & 393   \\
                        & Avg. block size (\%)  & 18.9  & ---   & ---   & ---  & 19.2  \\
                        & Matches               & 23    & ---   & ---   & ---  & 50    \\
                        \cmidrule(l){2-7} 
\multirow{3}{*}{\rotatebox[origin=c]{90}{\footnotesize sdhash}}
                        & Avg. block size (KB)  & 730   & 501   & 383   & 188  & 17.9  \\
                        & Avg. block size (\%)  & 35.7  & 24.5  & 18.7  & 9.17 & 0.88  \\
                        & Matches               & 34    & 44    & 50    & 50   & 50    \\
                        \cmidrule(l){2-7} 
\multirow{3}{*}{\rotatebox[origin=c]{90}{\footnotesize LZJD}}
                        & Avg. block size (KB)  & ---   & 868   & 376   & 16   & ---   \\
                        & Avg. block size (\%)  & ---   & 42.4  & 18.4  & 0.78 & ---   \\
                        & Matches               & ---   & 46    & 50    & 50   & ---   \\ \bottomrule
\end{tabular}
\end{adjustbox}
\end{table}

Inspecting the results for a 2MB total block size in \autoref{tbl:scb_2mb}, LZJD does not do well in this particular test. LZJD is unable to produce scores in the same large ranges as sdhash and ssdeep, but LZJD is also not designed to produce such scores. The use of completely random byte strings as the filler content of the SCB test also deflates the score LZJD gives, due to the increased number of sub-strings the Lempel-Ziv algorithm will find within these high entropy regions. This is a worst-case scenario for LZJD, as was theoretically analyzed in \cite{raff_lzjd_2017}. 

\begin{table}[!h]
\centering
\caption{Single Common Block results for a 512 KB file. Columns show the scores achieved, and rows the size of the common block of content, and the number of successful matches (max of 50). }
\label{tbl:scb_512kb}
\begin{adjustbox}{max width=\linewidth}
\begin{tabular}{@{}clccccc@{}}
\toprule
                        & Score                 & $\geq$25 & $\geq$20  & $\geq$15  & $\geq$10 & 0     \\ \midrule
\multirow{3}{*}{\rotatebox[origin=c]{90}{\footnotesize ssdeep}}
                        & Avg. block size (KB)  & 94.3  & 106   & ---   & ---  & 393   \\
                        & Avg. block size (\%)  & 18.4  & 20.6  & ---   & ---  & 19.2  \\
                        & Matches               & 28    & 5     & ---   & ---  & 50    \\
                        \cmidrule(l){2-7} 
\multirow{3}{*}{\rotatebox[origin=c]{90}{\footnotesize sdhash}}
                        & Avg. block size (KB)  & 185   & 160   & 140   & 107  & 16  \\
                        & Avg. block size (\%)  & 36.1  & 31.3  & 27.3  & 20.9 & 3.12  \\
                        & Matches               & 32    & 37    & 44    & 50   & 50    \\
                        \cmidrule(l){2-7} 
\multirow{3}{*}{\rotatebox[origin=c]{90}{\footnotesize LZJD}}
                        & Avg. block size (KB)  & 226   & 80.6  & 16.6  & ---  & ---   \\
                        & Avg. block size (\%)  & 44.2  & 15.8  & 3.25  & ---  & ---   \\
                        & Matches               & 39    & 50    & 50    & ---  & ---   \\ \bottomrule
\end{tabular}
\end{adjustbox}
\end{table}

The particular performance of LZJD at the lowest end of the score range is comparable to or better than sdhash, depending on which results are inspected. This can be better seen for a 512KB total block size, as shown in \autoref{tbl:scb_512kb}. Here we can see for a score of $\geq$20, sdhash requires a common block that is 31\% of the total block size, where LZJD requires a common block size of only 16\%. The results for the 8 MB total block size follow this pattern, and can be found in \ref{sec:full_frash}. 

We again note that this test is comparing the score of only two files at a time, and is not as relevant for LZJD since it does not try to produce the same types of score values as sdhash and ssdeep. LZJD's score is best interpreted as an approximate measure of the byte similarity of two files, and in practice, we will see that it is best viewed as an approximate lower bound on the percentage of similar bytes. 

Despite the SCB tests being a weak area for LZJD, the use of random bytes in the test construction also make this a worst-case scenario for LZJD. In practice, few files will make use of purely random byte sequences (which would have a byte entropy near 8). One of the only scenarios where we would expect the find such high-entropy sub-strings in a file is when dealing with malware and packed or compressed binaries, which corresponds to the scenario where LZJD was originally demonstrated to perform well \cite{raff_lzjd_2017}, where it was tested with Windows malware and Android APKs (which are compressed zip files). Still, removing the impact of completely random sub-strings on LZJD is an area for future research and improvement. 

\subsubsection{Random-noise-resistance} \label{sec:rand_resist}

The random noise test attempts to produce false negatives by randomly altering the file one byte at a time. After modification, the test records how many matches are achieved at each score and how many edits where required to reduce the score to that level. Bytes are altered via random insertions, deletions, and substitutions, and location is selected randomly. 

As noted in the original FRASH paper \cite{Breitinger2013}, the random noise resistance test is computationally demanding, and so we use only a random sample of 100 files from the t5 corpus. Our results find that LZJD is significantly more resistant to such alterations than either ssdeep or sdhash, which further increases the time it takes the tests to run. To reduce test runtime, after 200 edits, we begin altering the files by 10 bytes at a time. Once we reach 2000 edits, we increase to 100 edits at a time, and so on. We also add an early termination after 80\% of the file is altered, due to the extreme ranges that LZJD achieves in matching. 

The results of running the random noise test are shown in \autoref{tbl:rand_noise}, where \textit{Matches} indicates how many files achieved a given match score, and \textit{Avg. changes} is the average amount of bytes that needed to be altered for this score to appear, as a percentage of that file's size. For example, ssdeep was able to get a score equal to or higher than 70 for only 88 of the 100 files tested. It only took changing 0.005\% of the byte contents of a file to lower the score of ssdeep to this level. The better an algorithm's resistance to noise, the more we should be able to alter a file and still obtain a relatively high score. Because ssdeep and sdhash desire to produce a maximal score for any match, we would want to see a maximally high matching score for any percentage of edits. Under the LZJD interpretation of content similarity, we want the matching score to be similar to the percent of byte alterations performed. That is to say, if 25\% of the bytes were altered in the target file, we want to see LZJD return a score of 75 (i.e., 100-25\% = 75). 

\begin{table}[!htbp]
\centering
\caption{Random Noise tests. Best average number of changes needed to reduce the matching score to a specific level is shown in \textbf{bold}. }
\label{tbl:rand_noise}
\begin{adjustbox}{max width=\linewidth}
\begin{tabular}{@{}clccccc@{}}
\toprule
                        & Score                 & $\geq$70 & $\geq$50  & $\geq$40  & $\geq$25 & $\geq$10     \\ \midrule
\multirow{2}{*}{\rotatebox[origin=c]{90}{\footnotesize ssdeep}}
                        & Avg. changes (\%)     & 0.0052& 0.0206& 0.0615& ---  & ---  \\
                        & Matches               & 88    & 18    & 7     & ---  & ---   \\
                        \cmidrule(l){2-7} 
\multirow{2}{*}{\rotatebox[origin=c]{90}{\footnotesize sdhash}}
                        & Avg. changes (\%)     & \textbf{0.1068}& 0.2775& 0.3940& 0.5492& 0.9739\\
                        & Matches               & 96    & 99    & 100   & 100   & 99    \\
                        \cmidrule(l){2-7} 
\multirow{2}{*}{\rotatebox[origin=c]{90}{\footnotesize LZJD}}
                        & Avg. changes (\%)     & 0.0238& \textbf{0.6061}& \textbf{1.967}& \textbf{10.99}& \textbf{48.63} \\
                        & Matches               & 75    & 96    & 99    & 100  & 77   \\ \bottomrule
\end{tabular}
\end{adjustbox}
\end{table}

In examining the full results (see \ref{sec:full_frash}), it is clear that sdhash performs best when we consider only the higher scores ($\geq 55$). It routinely obtains the lowest percentage of average changes, followed by LZJD, and then ssdeep. While ssdeep is the only method to obtain the most high scores ($\geq 80$), this is of little utility due to the small number of changes needed to reduce such scores.

The robustness of LZJD becomes more apparent when we consider a score of $\geq$ 50, at which point LZJD requires twice as many byte edits to produce such a score compared to sdhash. Reducing LZJD to a score of $\geq 40$ required altering 1.97\% of the file, where sdhash produces a score of zero (no match) after an average of only 1.56\% of the file is edited. The rate at which LZJD's score is lowered decreases with each byte edit, and so its performance advantage improves dramatically relative to sdhash and ssdeep as we move down in matching score. Reducing LZJD to a score of 25 required 11.0\% of the bytes to be altered, which is 20 times greater than for sdhash. At the extreme end, reducing LZJD to a score of $\geq 10$ requires editing almost half the file. The 77 matches at this level is lower than 100 because the random noise test \textit{couldn't get LZJD to produce a score that low for many files}, and the FRASH test framework didn't anticipate a scenario where a score of 0 could not be obtained. This indicates a strong matching ability beyond the expectations of the FRASH designers. The FRASH code failed to count the files which obtained a score in the $(25, 10)$ range, and could not be reduced to the $[10, 0)$ before the test was forced to finish running by our modifications.

\subsubsection{Fragment Test} \label{sec:frag}

In the fragment tests of FRASH, a portion of the each file is removed, and then the remaining fragment is searched for against the database of all complete file hashes. The size of the fragment starts at 95\%, nearly the whole file, and decreases down to only a 1\% portion of the original file. The motivation of these tests are to determine how small a fragment can be while still being matched with the source file. This scenario may occur with any storage or transport format where a file may be broken up into chunks, such as the fragment storage in a file system or individual packets in network traffic. 

FRASH runs these fragmentation tests in two modes, one where the file has data removed from the end only (\textit{end cut}), and one where a random portion of the file is removed from both the beginning and end of the file (\textit{random cut}). In the former case, the fragment always starts as the same string of bytes but ends prematurely. In the latter case, the fragment is essentially a random portion of the file (and most likely from near the middle of the original file). 
The results of the fragment tests are presented in the next two tables. In each table, the File Size (\%) is the size of the file fragment as a percentage of the original file it came from. 

\begin{table}[!htbp]
\centering
\caption{Fragment detection test result, random cut. Column indicates size of the fragment with respect to the source file. Rows show percent of correctly matched files and average score for correctly matched files.}
\label{tbl:frag_randStart}
\begin{adjustbox}{max width=\linewidth}
\begin{tabular}{@{}clcccccc@{}}
\toprule
\multicolumn{2}{c}{File Size (\%)} & 95           & 50           & 10                        & 5             & 3             & 1             \\ \midrule
\multirow{2}{*}{\rotatebox[origin=c]{90}{\footnotesize ssdeep}}
              & Matches (\%)       & 99.9         & 91.3         & 0.65                      & \textless0.01 & 0             & 0             \\
              & Avg. Score         & 96.7         & 65.9         & 46.2                      & 61.0          & ---           & ---           \\
\addlinespace[0.4em]
\multirow{2}{*}{\rotatebox[origin=c]{90}{\footnotesize sdhash}}
              & Matches (\%)       & \textbf{100} & \textbf{100} & 98.1                      & 90.6          & 81.1          & 57.9          \\
              & Avg. Score         & 83.4         & 68.5         & 75.7                      & 73.4          & 76.7          & 81.0          \\
\addlinespace[0.4em]
\multirow{2}{*}{\rotatebox[origin=c]{90}{\footnotesize LZJD}}
              & Matches (\%)       & \textbf{100} & \textbf{100} & \textbf{\textgreater99.9} & \textbf{99.9} & \textbf{99.4} & \textbf{98.5} \\
              & Avg. Score         & 72.4         & 24.9         & 6.43                      & 3.88          & 2.73          & 1.31          \\ 
\bottomrule
\end{tabular}
\end{adjustbox}
\end{table}

\begin{table}[!htbp]
\centering
\caption{Fragment detection test result, end cut. Column indicates size of the fragment with respect to the source file. Rows show percent of correctly matched files and average score for correctly matched files. }
\label{tbl:frag_right}
\begin{adjustbox}{max width=\linewidth}
\begin{tabular}{@{}clcccccc@{}}
\toprule
\multicolumn{2}{c}{File Size (\%)} & 95           & 50           & 10           & 5            & 3            & 1             \\ \midrule
\multirow{2}{*}{\rotatebox[origin=c]{90}{\footnotesize ssdeep}}
              & Matches (\%)       & \textbf{100} & 93.1         & 1.73         & 0.49         & 0.20         & 0             \\
              & Avg. Score         & 97.7         & 71.7         & 56.9         & 55.7         & 47.9         & ---           \\
\addlinespace[0.4em]
\multirow{2}{*}{\rotatebox[origin=c]{90}{\footnotesize sdhash}}
              & Matches (\%)       & \textbf{100} & \textbf{100} & 98.3         & 91.1         & 82.5         & 58.7          \\
              & Avg. Score         & 97.3         & 99.5         & 97.9         & 96.9         & 95.04        & 90.5          \\
\addlinespace[0.4em]
\multirow{2}{*}{\rotatebox[origin=c]{90}{\footnotesize LZJD}}
              & Matches (\%)       & \textbf{100} & \textbf{100} & \textbf{100} & \textbf{100} & \textbf{100} & \textbf{99.7} \\
              & Avg. Score         & 92.8         & 40.1         & 8.33         & 4.63         & 3.09         & 1.36          \\ \bottomrule
\end{tabular}
\end{adjustbox}
\end{table}

The ssdeep algorithm is particularly vulnerable to this approach, and is significantly degraded in its ability to correctly match files by a fragment being just 50\% of the original file size. Sdhash is more robust, and is not meaningfully impacted in matching ability until fragments are 5\% of the original file size or less, where it starts to quickly degrade in accuracy. We also notice a confusing behavior in the average matched score produced by sdhash. In \autoref{tbl:frag_right}, the sdhash score slowly decreases from high 90s to low 90s, which is a reasonable behavior to expect as the fragment size decreases. However, in \autoref{tbl:frag_randStart}, the sdhash score first decreases from the low 80s to the low 70s, and then begins increasing against back into the low 80s. 

Compared to sdhash, LZJD obtains lower average matching scores. In \autoref{tbl:frag_right}, these scores are nearly perfectly aligned with the interpretation of a similarity of X\% indicating that the X\% of the contents are the same. The scores returned for LZJD are a bit below this expectation in \autoref{tbl:frag_randStart}, but still match the general trend. 
This can be explained by the LZSet construction process being sensitive to changes in the byte string, causing changes in the set. 
In the case of \autoref{tbl:frag_right}, corresponding to the end cut version of the fragment test, the start of the byte string will remain unchanged. This means the LZ set generated will also be generated in the same order, and will simply stop early once the fragment comes to a premature end. This results in a high quality match of LZ set contents when computing the Jaccard similarity. In the random cut case of \autoref{tbl:frag_randStart}, the beginning of the file has been removed. This changes the set of sub-strings computed by the LZ approach, resulting in a lowered match. However the match is still robust, as evident by the high number of matches LZJD obtained. 

\begin{figure}
\centering
\begin{tikzpicture}
\begin{axis}[
	legend pos=outer south,
    legend columns=3, 
    xlabel=Fragment Size Percentage,
    ylabel=Number Files Matched,
    ]
    \addplot[dashed,green,mark=square,mark options={solid}] table [x=Size, y=ssdeep_matches, col sep=comma] {csvs/frag_random.csv};
    \addplot[dashed,blue,mark=o,mark options={solid}] table [x=Size, y=sdhash_matches, col sep=comma] {csvs/frag_random.csv};
    \addplot[red,mark=x] table [x=Size, y=lzjd_matches, col sep=comma] {csvs/frag_random.csv};
    \legend{ssdeep,sdhash,LZJD}
    
\end{axis}
\end{tikzpicture}
\caption{Fragment detection random-cut results, x-axis shows the fragment size as a precentage of the original file, and y-axis shows the number of files correctly matched. }
\label{fig:frag_random}
\end{figure}
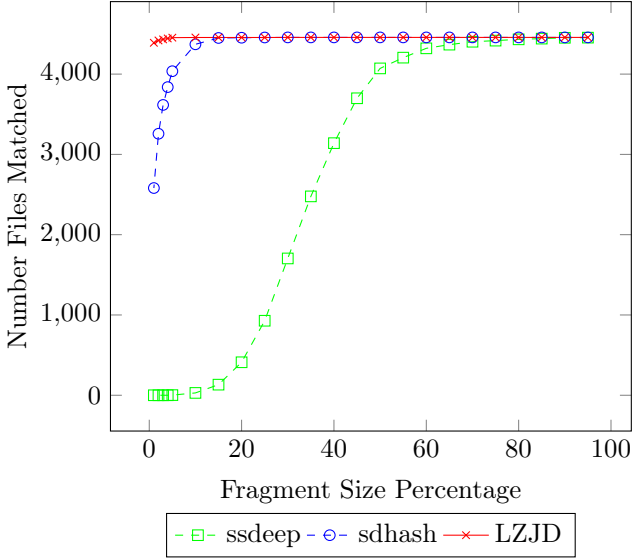

This robustness in matching ability is emphasized in \autoref{fig:frag_random}, where we plot the number of correct matches in the random-cut test against the size of the fragment as a percentage of the original file. We can clearly see ssdeep requires fragments to be 60\% of the original file or larger to get reliable matches. Sdhash holds for a larger range, but begins dropping once the fragments are 10\% of the original file or less. LZJD performs well across all sizes, still obtaining the majority of matches even at 1\% size. The end-cut version of the fragmentation tests are similar, and can be numerically compared in the Appendix. 

We claim that this robustness is the more important property. The fragment results  support the conclusion that LZJD is more robust in its ability to match small fragments to their source files compared to ssdeep and sdhash. In all cases, LZJD is either tied with or better than sdhash at this task. Even down to 1\% fragment sizes, LZJD is able to match 99\% of fragments to their source file. In comparison, sdhash is only able to match just under 60\% of fragments.

\subsubsection{Alignment Test} \label{sec:align}

One area of weakness for many similarity hash functions is padding inserted at the beginning of a file.  Ssdeep in particular is weak in this scenario \cite{Baier2011}. The alignment test in FRASH is designed for this scenario, and inserts random bytes into the beginning of a file, and then attempts to match it back against the full database. An analysis of the LZSet algorithm used by LZJD may also lead one to assume that LZJD is susceptible to this same problem. Because the LZSet is built incrementally, strings seen earlier can impact the LZSet, changing what is captured in the later sections of the byte string. The results of this section will show that while this could be a problem for LZJD in the limit, the performance on the FRASH tests indicate that its matching ability is not hampered by this scenario.

The FRASH tests for matching in-spite of excess padding is run in two modes. One where a fixed number of bytes are added to the file, and the other where a fixed percentage of the original file size is added to the front. The results for the latter scenario are presented in \autoref{tbl:alignment_percentage}. We present only the percentage results as they are the most aggressive and challenging version of the test. 

\begin{table}[!htbp]
\centering
\caption{Alignment test result. Column shows the size of the added bytes, as a percentage of original file size. Rows show percent of correctly matched files and average score for correctly matched files. }
\label{tbl:alignment_percentage}
\begin{tabular}{@{}clccccc@{}}
\toprule
\multicolumn{2}{c}{Added (\%)} & 10           & 50           & 100          & 300          & 500          \\ \midrule
\multirow{2}{*}{\rotatebox[origin=c]{90}{\footnotesize ssdeep}}
            & Matches (\%)     & 99.6         & 92.8         & 72.4         & 3.03         & 0            \\
            & Avg. Score       & 91.1         & 71.7         & 60.1         & 35.3         & ---          \\
\addlinespace[0.4em]
\multirow{2}{*}{\rotatebox[origin=c]{90}{\footnotesize sdhash}}
            & Matches (\%)     & \textbf{100} & \textbf{100} & \textbf{100} & \textbf{100} & \textbf{100} \\
            & Avg. Score       & 67.6         & 69.2         & 68.6         & 68.8         & 68.2         \\
\addlinespace[0.4em]
\multirow{2}{*}{\rotatebox[origin=c]{90}{\footnotesize LZJD}}
            & Matches (\%)     & \textbf{100} & \textbf{100} & \textbf{100} & \textbf{100} & \textbf{100} \\
            & Avg. Score       & 40.9         & 22.1         & 14.8         & 6.79         & 4.53         \\ \bottomrule
\end{tabular}
\end{table}

As expected, we can see that ssdeep is significantly impacted by the front-padding of the binary, and can only match 3\% of files when 300\% of the file size is padded to the front. Both sdhash and LZJD are able to match 100\% of files in the tested range. We also see that the scores for both are negatively impacted by the addition of the bytes to the front of the file. For sdhash, the scores are in the high 60s instead of the normal 80s-90s that it is able to achieve in the other benchmarks. Because there is no particular interpretation that applies to the sdhash score, we can not offer any analysis as to cause or reason. 

For LZJD, we would expect a score in the range of $1/(1+x/100)$, where $x$ is the percent of the file size added as padding. In each case, the LZJD score is one third to one half of this expected value. This can be explained by the Lempel-Ziv encoding scheme, which creates a maximal number of entries in the set when presented with high entropy (i.e., random looking) data. Because the x\% of bytes added by FRASH are random, this will create disproportionately more entries in the LZ set, and thus become a larger portion of the hash digest. The effect is that there will be considerably more than x\% new hashes added to the set, with the amount more being dependent upon the normal entropy of the file under consideration. Because these entries in the hash are from random sub-strings, they are unlikely to appear in another file, and so they are not matched and the score is reduced.

\section{Discussion} \label{sec:discussion}

At this point we have performed extensive testing of LZJD compared to ssdeep and sdhash. It is faster to hash, faster at hash comparisons, produces more compact hashes, and provides higher matching accuracy for smaller files, compared to these previous tools. Only ssdeep is faster at hashing and has smaller digests, but its matching ability is not sufficient for the multitude of file types in the t5 corpus. This coalesces to a strong argument for the use of LZJD as an alternative to ssdeep and sdhash for digital forensic applications. The faster comparison time and accuracy combined will allow LZJD to be used in real deployments with databases larger than what either ssdeep or sdhash can handle, while stemming a natural increase in false positives due to the use of larger datasets. This runtime advantage is critical for tool adoption, as practitioners would be unlikely to make use of a tool that did not produce timely results. 

\subsection{LZJD use Compared to Ssdeep and Sdhash} \label{sec:lzjd_cpm_others}
\todo[inline]{Below paragraphs are new / replacements. Good? }

The most significant difference between LZJD and prior similarity digests is the nature of the score value produced. LZJD, like ssdeep and sdhash, will need a "significance" threshold to be determined which may change for different file types and scenarios. The difference comes in the nature of the score's value itself. For ssdeep and sdhash, the exact score $x$ has no particular meaning, and instead a single meaning is often prescribed to only a few ranges of values. For example, the $[21-100]$, $[11-20]$, and $[1-10]$ recommended for sdhash divide up the entire positive range of values into classifications of "Strong", "Marginal", and "Weak" correlation \cite{Roussev2012}.  This can be uninformative when multiple files produce high scores --- an issue that occurred in our malware experiments in \autoref{sec:lzjd_speedup_results}. 

For LZJD, we can interpret the score as a rough measure of byte similarity, or more precisely, as an approximate lower bound on a normalized edit-distance between the files. Not only does this give us an interpretation of the score returned by LZJD, but we can use it to infer what a reasonable threshold might be for many file types and scenarios. This may require more thought on the practitioner's part, that is to make an estimate of what the expected overlap between files might be, or what the maximum score one might expect. Though this requires more mental effort, it is not a requirement --- and users could choose to empirically determine their desired scores just as they have done with ssdeep and sdhash. We believe that this interpretation though will ultimately aid its use not just by giving it meaning, but avoiding failure-cases that can occur without such a background (as exemplified by sdhash's failure in \autoref{sec:lzjd_speedup_results}). 

To give concrete examples of what we mean, consider that file types such as PDFs and EXEs have some amount of boiler-plate mandated by the file format's specification, or may simply be common to most files of that type. In this scenario, one would expect LZJD to produce a minimal score dependent on how much of the boilerplate or common content is shared across files. If the practitioner knows what this level of boilerplate is, they can use that as a minimum-threshold for potential matches. 

As another example, consider the results of the fragment tests in \autoref{sec:frag}. 
If an analyst were to use LZJD in this fragment scenario, where it is known that we have a $\alpha$ byte long file fragment that we want to compare against a known (larger) file of length $\beta$, it may then be reasonable to use an adjusted scoring of $\text{sim}(\alpha, \beta)\cdot \beta / \alpha$ to adjust for the fact that our expected similarity should not generally exceed the ratio of the differences in file length (i.e., $\alpha/\beta$). This requires thought on the analyst's part to realize that smaller scores should be expected, but accurate matching is still possible --- and thus might want to alter their interpretation of the score's significance. 

\subsection{LZJD Score Interpretation} \label{sec:lzjd_score_interpretation}

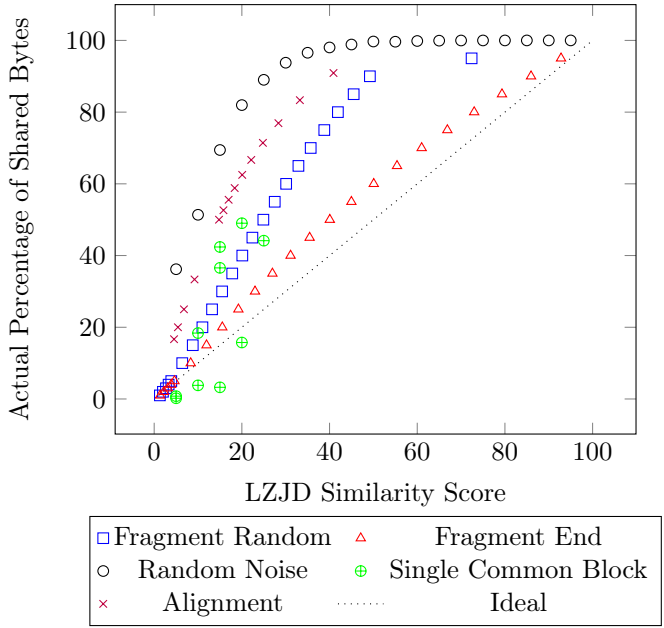
\begin{figure}
\centering
\begin{tikzpicture}
\begin{axis}[
	legend pos=outer south,
    legend columns=2, 
    xlabel=LZJD Similarity Score,
    ylabel=Actual Percentage of Shared Bytes
    ]
    \addplot[
        scatter,only marks,scatter src=explicit symbolic,
        scatter/classes={
            a={mark=square,blue},
            b={mark=triangle,red},
            c={mark=o,draw=black,fill=black},
            d={mark=oplus,green},
            e={mark=x,purple}
        }
    ]
    table [x=actual, y=expected, meta=label, col sep=comma] {lzjd_lower_bound_emperical.csv};
    \addplot[dotted, mark=none, black, samples=2] coordinates {(1.0,1.0) (100,100.0)};
    \legend{Fragment Random,Fragment End,Random Noise,Single Common Block,Alignment,Ideal}
    
\end{axis}
\end{tikzpicture}
\caption{Comparing the LZJD similarity score (x-axis) with the actual percentage of altered or added bytes (y-axis) for all tests run by the FRASH suite. The ideal 1-to-1 correspondence \eqref{eq:lzjd_lb} is shown as a dashed line. Values above this line indicate LZJD under-estimating the change in bytes. }
\label{fig:lzjd_vs_actual_change}
\end{figure}

\todo[inline]{Below paragraphs adjusted, Good? }
As we have discussed throughout this work, LZJD's score is more interpretable than the ones returned by ssdeep and sdhash. 
We noted in \autoref{sec:background_lzjd} that LZJD's score can be loosely interpreted as the percentage of shared byte contents between files, and empirically tends to act as a lower bound on a normalized edit distance (as specified in equation \eqref{eq:lzjd_lb}). 
The extensive experiments provided by FRASH in \autoref{sec:frash_lzjd}  support this conclusion. To condense these results, we plot in \autoref{fig:lzjd_vs_actual_change} the relationship between LZJD's score and actual percentage of shared bytes. This percentage can be determined for all of the FRASH tests, though would not be known {\it a priori} in practice. These results show LZJD almost uniformly under-estimating the percentage of bytes altered. The only exception being the five points from the Single Common Block tests, three of which are from the most-extreme terminating state. This overall result leads us to recommend treating LZJD's score as a \textit{lower bound} on the percentage of bytes altered. That is to say, if LZJD returns a score of 23, then it is relatively safe to assume that the two files share at least 23\% of their byte contents with each other. While this is not a guarantee, it is empirically supported by a considerable majority of test cases (84 out of 89 data points) and we believe will be useful to the practitioner. 

The immediate question would be why does LZJD tend to produce a lower bound estimate? 
The LZSet method that produces the initial set of sub-strings is sensitive to single byte alterations. Because the set is constructed in a sequential manner, once one byte is altered, it has the potential to propagate forward and alter the rest of the set. This byte sensitivity is what causes LZJD to act as a lower bound, and is the reason why it is often difficult for LZJD to obtain high match scores above 50\%. Despite this weakness LZJD operates effectively, and the few cases where LZJD over-estimated the percentage of bytes changed are cases where LZJD successfully matched 100\% of the altered files to their original sources. 

This is also connected with the effect of random byte sequences on LZJD's similarity score, as random bytes will cause the same impact on the LZSet. The impact of random bytes is tested by the SCB, Alignment, and Random Noise tests in FRASH. All three of theses tests make use of random bytes to create the test case files. These tests show that LZJD can perform well even if random bytes are present, but does tend to impact the similarity score LZJD returns. The Malware tests in \autoref{sec:lzjd_speedup_results} also test a higher average entropy file than the t5 corpus, which has become the standard benchmark corpus for similarity digests. The exact entropy statistics are shown in \autoref{tbl:corpus_entropy}. The performance of LZJD in accurately matching nearest neighbors when the average and median file entropy is as high as 7.96 shows that this weakness does not halt LZJD's matching ability. 

\begin{table}[!htb]
\centering
\caption{Statistics on file entropy broken down by each corpus used in this work. }
\label{tbl:corpus_entropy}
\begin{tabular}{@{}lccccc@{}}
\toprule
        & \multicolumn{2}{c}{Kaggle}       & \multicolumn{2}{c}{Android} &      \\
\cmidrule(lr){2-3} \cmidrule(lr){4-5} 
Entropy & Bytes            & ASM           & APK          & TAR          & t5   \\ 
\midrule
Average & 6.73            & 4.48           & 7.96         & 6.68         & 5.88 \\
Median  & 6.83            & 4.51           & 7.96         & 6.77         & 5.32 \\
Min     & 1.64            & 3.83           & 4.10         & 2.61         & 0.21 \\
Max     & 7.85            & 5.35           & 8.00         & 8.00         & 8.00 \\ \bottomrule
\end{tabular}
\end{table}

Since current methods used for digest similarity do not return interpretable similarity scores, there are no current use-cases to compare LZJD against. As analysts begin to use LZJD, we believe the interpretability will become useful to practitioners. \todo[inline]{wording?}Investigating the reality of these hypotheses is beyond the scope of this work.  In particular, this new ability may have an impact on:
\begin{enumerate}
\item New user training. Being able to explain the results that a method produces is a natural way to help new users learn and understand their tools, and the LZJD algorithm itself can be specified with only a few lines of code. This may aid in helping and enabling tool adoption to a wider breadth of professionals and skill sets. 
\item Evidence and testimony. In legal proceedings there is often a need to present evidence to support a case, either in court or in the pursuit of an arrest warrant, for example. That LZJD can be described in a less technical manner as a "conservative estimate of shared content" between two files could be useful in this regard, and is empirically supported. The  exact interpretation as the intersection of compression dictionaries is available as well for more technical needs. 
\item Machine Learning applications. While ssdeep and sdhash have been used with other machine learning methods before, they both lack the nice metric space and kernel properties of the Jaccard distance that LZJD inherits. We suspect LZJD will thus find wider user with machine learning methods, and its interpretable descriptions will aid in being able to explain and interpret larger models built using LZJD as a component. 
\end{enumerate}

To our knowledge, there has yet to be any discussion on what the ideal scoring approach would be for a similarity digest. Our results open an opportunity to discuss such potential design choices. In particular, should scores indicate a level of similarity (resemblance), or a level of commonality (containment)? By this we mean, should scores be interpretable as a measure of how much content of two files are \textit{shared} in aggregate (as LZJD currently does)? Or should scores reflect that two byte strings share some commonality, such as being from the same file, or how much one file could be \textit{subsumed} by another (as sdhash does)? For LZJD, we have already given one instance in which its design could be modified to reflect a preference for commonality when searching for the source of a file fragment (see the discussion near the end of \autoref{sec:lzjd_cpm_others}). There may also be other goals toward which one could design a similarity digest, but leave further discussion of this question for future work. 

\subsubsection{On Resemblance and Containment}

\todo[inline]{New. Good?}
We take a moment to further discuss the resemblance vs containment question with respect to the results we saw with LZJD. As mentioned in \autoref{sec:background_lzjd}, LZJD measures the resemblance between LZ sets $A$ and $B$. That we use the Jaccard similarity, for the purpose of computing resemblance, is what allows us to develop a digest of fixed size. Another potential measure of interest is containment, which can be expressed as \eqref{eq:containment}. 

\begin{equation}\label{eq:containment}
c(A, B) = \frac{|A \cap B |}{|A|}
\end{equation}

Containment asks how much of set $A$ is contained within set $B$. Sdhash's variable length digest sizes allow it to answer queries regarding either containment or resemblance fashion \cite{Roussev2012}. Answering containment queries in an unbiased manner requires such variable-length digests \cite{Broder:1997:RCD:829502.830043}. 

The FRASH Fragmentation, Alignment, and Single Common Block (SCB) tests (\autoref{sec:frag}, \autoref{sec:align} and \autoref{sec:scb} respectively) are tests of containment. LZJD out-performs both ssdeep and sdhash in the fragmentation tests, especially for extreme cases. LZJD ties with sdhash in obtaining all matches for Alignment, and LZJD has only comparable performance to sdhash in the SCB tests. One may then wonder, if LZJD is answering resemblance, how is it able to do well at these containment tasks, and even outperform approaches that should have an advantage? 

We believe the insight into understanding this approach is to recognize that $c(A, B) \geq J(A, B) \geq 0$. That is to say, the resemblance query is necessarily a lower bound on containment. If the correct containment score is zero, the resemblance score must necessarily also be zero. Thus LZJD will never over-estimate the containment case. Obtaining the same matching scores then relies on obtaining the same rank ordering between resemblance and containment. Our results with the FRASH tests would indicate that LZJD does well in this regard, as it achieves matching performance comparable to or better than sdhash in all tests. 

\subsection{Future Work}

Another advantage of the LZJD approach, which we have not tested in this work, is further scaling abilities of the digest hash. Because the LZJD hash produces a valid distance metric, it is possible to use metric indexes to prune distance computations from a search\cite{Beygelzimer2006,Yianilos1993}. Further speedups can be obtained by performing partial digest comparisons. Because the LZJD hash is obtained by selecting the $k$ smallest hash values, every LZJD digest of length $k$ contains the $k'$ digest $\forall k' < k$. This gives a natural way to balance between speed and accuracy. We leave exploring these options to future work. 

File size is also an important consideration in digest construction and application. This has been tested to some degree by the FRASH suite and our Malware classification tests. The fragment tests in section \autoref{sec:frag} are explicitly testing matching performance when file sizes differ by up to two orders of magnitude (the original file compared to a 1\% fragment).  The Kaggle ASM corpus has an average file size of 13.5 MB, compared to only 425 KB for the t5 corpus normally used. In both of these cases, LZJD outperforms ssdeep and sdhash by wide margins. Further exploring the impact of large file size comparisons (GB vs GB) and disparate size comparisons (GB vs KB) is an important topic. In particular, what files should be used, and what are the real-life scenarios that should be simulated? 

\section{Conclusions} \label{sec:conclusion}

The Lempel-Ziv Jaccard Distance was introduced to address problems in malware classification, but we have shown that it has significant utility as a similarity digest for digital forensic applications. Compared to existing tools, such as sdhash, LZJD offers a non-heuristic score that can be interpreted by the user as the amount of byte similarity between two files. Beyond this property, LZJD is more robust in its ability to match file fragments to their source, even when forced to match a fragment on the order of 1\% of the original file's size.  We have also shown that LZJD can be made practical from a speed perspective, with digest comparison over 60 times faster than sdhash's, and hashing time 34\% faster. This will allow the use of larger search databases than is possible with other tools, while also being more accurate. In the interest of tool adoption, we have released an open-source implementation that mimics sdhash's command line options. This should allow LZJD to be easily integrated with existing work-flows for fast adoption by practitioners. 

\section*{References}

\Urlmuskip=0mu plus 1mu\relax
\bibliography{Mendeley}

\begin{thebibliography}{42}
\providecommand{\natexlab}[1]{#1}
\providecommand{\url}[1]{\texttt{#1}}
\providecommand{\urlprefix}{URL }
\expandafter\ifx\csname urlstyle\endcsname\relax
  \providecommand{\doi}[1]{doi:\discretionary{}{}{}#1}\else
  \providecommand{\doi}[1]{doi:\discretionary{}{}{}\begingroup
  \urlstyle{rm}\url{#1}\endgroup}\fi
\providecommand{\bibinfo}[2]{#2}

\bibitem[{Harichandran et~al.(2016)Harichandran, Breitinger, and
  Baggili}]{Harichandran2016}
\bibinfo{author}{V.~Harichandran}, \bibinfo{author}{F.~Breitinger},
  \bibinfo{author}{I.~Baggili}, \bibinfo{title}{{Bytewise Approximate Matching:
  The Good, The Bad, and The Unknown}}, \bibinfo{journal}{Journal of Digital
  Forensics, Security and Law} \bibinfo{volume}{11}~(\bibinfo{number}{2})
  (\bibinfo{year}{2016}) \bibinfo{pages}{59--78}, ISSN
  \bibinfo{issn}{15587223}, \doi{\bibinfo{doi}{10.15394/jdfsl.2016.1379}},
  \urlprefix\url{http://commons.erau.edu/jdfsl/vol11/iss2/4/}.

\bibitem[{Roussev and Quates(2012)}]{Roussev2012}
\bibinfo{author}{V.~Roussev}, \bibinfo{author}{C.~Quates},
  \bibinfo{title}{{Content triage with similarity digests: The M57 case
  study}}, \bibinfo{journal}{Digital Investigation} \bibinfo{volume}{9}
  (\bibinfo{year}{2012}) \bibinfo{pages}{S60--S68}, ISSN
  \bibinfo{issn}{17422876}, \doi{\bibinfo{doi}{10.1016/j.diin.2012.05.012}}.

\bibitem[{Costin et~al.(2014)Costin, Zaddach, Francillon, and
  Balzarotti}]{Costin:2014:LAS:2671225.2671232}
\bibinfo{author}{A.~Costin}, \bibinfo{author}{J.~Zaddach},
  \bibinfo{author}{A.~Francillon}, \bibinfo{author}{D.~Balzarotti},
  \bibinfo{title}{{A Large-scale Analysis of the Security of Embedded
  Firmwares}}, in: \bibinfo{booktitle}{Proceedings of the 23rd USENIX
  Conference on Security Symposium}, SEC'14, \bibinfo{publisher}{USENIX
  Association}, \bibinfo{address}{Berkeley, CA, USA}, ISBN
  \bibinfo{isbn}{978-1-931971-15-7}, \bibinfo{pages}{95--110},
  \urlprefix\url{http://dl.acm.org/citation.cfm?id=2671225.2671232},
  \bibinfo{year}{2014}.

\bibitem[{Jang et~al.(2011)Jang, Brumley, and Venkataraman}]{Jang2011}
\bibinfo{author}{J.~Jang}, \bibinfo{author}{D.~Brumley},
  \bibinfo{author}{S.~Venkataraman}, \bibinfo{title}{{BitShred: Feature Hashing
  Malware for Scalable Triage and Semantic Analysis}}, in:
  \bibinfo{booktitle}{Proceedings of the 18th ACM conference on Computer and
  communications security - CCS}, \bibinfo{publisher}{ACM Press},
  \bibinfo{address}{New York, New York, USA}, ISBN
  \bibinfo{isbn}{9781450309486}, \bibinfo{pages}{309--320},
  \doi{\bibinfo{doi}{10.1145/2046707.2046742}},
  \urlprefix\url{http://dl.acm.org/citation.cfm?doid=2046707.2046742},
  \bibinfo{year}{2011}.

\bibitem[{Lakhotia et~al.(2013)Lakhotia, Walenstein, Miles, and
  Singh}]{Lakhotia:2013:VRL:2509225.2509228}
\bibinfo{author}{A.~Lakhotia}, \bibinfo{author}{A.~Walenstein},
  \bibinfo{author}{C.~Miles}, \bibinfo{author}{A.~Singh},
  \bibinfo{title}{{VILO: A Rapid Learning Nearest-neighbor Classifier for
  Malware Triage}}, \bibinfo{journal}{Journal in Computer Virology}
  \bibinfo{volume}{9}~(\bibinfo{number}{3}) (\bibinfo{year}{2013})
  \bibinfo{pages}{109--123}, ISSN \bibinfo{issn}{1772-9890},
  \doi{\bibinfo{doi}{10.1007/s11416-013-0178-3}},
  \urlprefix\url{http://dx.doi.org/10.1007/s11416-013-0178-3}.

\bibitem[{Breitinger et~al.(2013{\natexlab{a}})Breitinger, Astebol, Baier, and
  Busch}]{Breitinger:2013:MNA:2496032.2497148}
\bibinfo{author}{F.~Breitinger}, \bibinfo{author}{K.~P. Astebol},
  \bibinfo{author}{H.~Baier}, \bibinfo{author}{C.~Busch},
  \bibinfo{title}{{mvHash-B - A New Approach for Similarity Preserving
  Hashing}}, in: \bibinfo{booktitle}{Proceedings of the 2013 Seventh
  International Conference on IT Security Incident Management and IT
  Forensics}, IMF '13, \bibinfo{publisher}{IEEE Computer Society},
  \bibinfo{address}{Washington, DC, USA}, ISBN
  \bibinfo{isbn}{978-0-7695-4955-2}, \bibinfo{pages}{33--44},
  \doi{\bibinfo{doi}{10.1109/IMF.2013.18}},
  \urlprefix\url{http://dx.doi.org/10.1109/IMF.2013.18},
  \bibinfo{year}{2013}{\natexlab{a}}.

\bibitem[{Breitinger and Baier(2013)}]{Breitinger2013a}
\bibinfo{author}{F.~Breitinger}, \bibinfo{author}{H.~Baier},
  \bibinfo{title}{{Similarity Preserving Hashing: Eligible Properties and a New
  Algorithm MRSH-v2}}, in: \bibinfo{booktitle}{Digital Forensics and Cyber
  Crime}, \bibinfo{pages}{167--182},
  \doi{\bibinfo{doi}{10.1007/978-3-642-39891-9{\_}11}},
  \urlprefix\url{http://link.springer.com/10.1007/978-3-642-39891-9_11},
  \bibinfo{year}{2013}.

\bibitem[{Winter et~al.(2013{\natexlab{a}})Winter, Schneider, and
  Yannikos}]{Winter2013}
\bibinfo{author}{C.~Winter}, \bibinfo{author}{M.~Schneider},
  \bibinfo{author}{Y.~Yannikos}, \bibinfo{title}{{F2S2: Fast forensic
  similarity search through indexing piecewise hash signatures}},
  \bibinfo{journal}{Digital Investigation}
  \bibinfo{volume}{10}~(\bibinfo{number}{4})
  (\bibinfo{year}{2013}{\natexlab{a}}) \bibinfo{pages}{361--371}, ISSN
  \bibinfo{issn}{17422876}, \doi{\bibinfo{doi}{10.1016/j.diin.2013.08.003}},
  \urlprefix\url{http://linkinghub.elsevier.com/retrieve/pii/S1742287613000789}.

\bibitem[{Kornblum(2006)}]{Kornblum2006}
\bibinfo{author}{J.~Kornblum}, \bibinfo{title}{{Identifying almost identical
  files using context triggered piecewise hashing}}, \bibinfo{journal}{Digital
  Investigation} \bibinfo{volume}{3} (\bibinfo{year}{2006})
  \bibinfo{pages}{91--97}, ISSN \bibinfo{issn}{17422876},
  \doi{\bibinfo{doi}{10.1016/j.diin.2006.06.015}}.

\bibitem[{Roussev(2011)}]{Roussev2011}
\bibinfo{author}{V.~Roussev}, \bibinfo{title}{{An evaluation of forensic
  similarity hashes}}, \bibinfo{journal}{Digital Investigation}
  \bibinfo{volume}{8} (\bibinfo{year}{2011}) \bibinfo{pages}{S34--S41}, ISSN
  \bibinfo{issn}{17422876}, \doi{\bibinfo{doi}{10.1016/j.diin.2011.05.005}}.

\bibitem[{Roussev(2010)}]{Roussev2010}
\bibinfo{author}{V.~Roussev}, \bibinfo{title}{{Data Fingerprinting with
  Similarity Digests}}, in: \bibinfo{editor}{K.-P. Chow},
  \bibinfo{editor}{S.~Shenoi} (Eds.), \bibinfo{booktitle}{Advances in Digital
  Forensics VI: Sixth IFIP WG 11.9 International Conference on Digital
  Forensics, Hong Kong, China, January 4-6, 2010, Revised Selected Papers},
  \bibinfo{publisher}{Springer Berlin Heidelberg}, \bibinfo{address}{Berlin,
  Heidelberg}, ISBN \bibinfo{isbn}{978-3-642-15506-2},
  \bibinfo{pages}{207--226},
  \doi{\bibinfo{doi}{10.1007/978-3-642-15506-2{\_}15}},
  \urlprefix\url{http://dx.doi.org/10.1007/978-3-642-15506-2_15},
  \bibinfo{year}{2010}.

\bibitem[{Li et~al.(2015)Li, Sundaramurthy, Bardas, Ou, Caragea, Hu, and
  Jang}]{191669}
\bibinfo{author}{Y.~Li}, \bibinfo{author}{S.~C. Sundaramurthy},
  \bibinfo{author}{A.~G. Bardas}, \bibinfo{author}{X.~Ou},
  \bibinfo{author}{D.~Caragea}, \bibinfo{author}{X.~Hu},
  \bibinfo{author}{J.~Jang}, \bibinfo{title}{{Experimental Study of Fuzzy
  Hashing in Malware Clustering Analysis}}, in: \bibinfo{booktitle}{8th
  Workshop on Cyber Security Experimentation and Test (CSET 15)},
  \bibinfo{publisher}{USENIX Association}, \bibinfo{address}{Washington, D.C.},
  \urlprefix\url{https://www.usenix.org/conference/cset15/workshop-program/presentation/li},
  \bibinfo{year}{2015}.

\bibitem[{Raff and Nicholas(2017)}]{raff_lzjd_2017}
\bibinfo{author}{E.~Raff}, \bibinfo{author}{C.~Nicholas}, \bibinfo{title}{{An
  Alternative to NCD for Large Sequences, Lempel-Ziv Jaccard Distance}}, in:
  \bibinfo{booktitle}{Proceedings of the 23rd ACM SIGKDD International
  Conference on Knowledge Discovery and Data Mining - KDD '17},
  \bibinfo{publisher}{ACM Press}, \bibinfo{address}{New York, New York, USA},
  ISBN \bibinfo{isbn}{9781450348874}, \bibinfo{pages}{1007--1015},
  \doi{\bibinfo{doi}{10.1145/3097983.3098111}},
  \urlprefix\url{http://dl.acm.org/citation.cfm?doid=3097983.3098111},
  \bibinfo{year}{2017}.

\bibitem[{Breitinger et~al.(2013{\natexlab{b}})Breitinger, Stivaktakis, and
  Baier}]{Breitinger2013}
\bibinfo{author}{F.~Breitinger}, \bibinfo{author}{G.~Stivaktakis},
  \bibinfo{author}{H.~Baier}, \bibinfo{title}{{FRASH: A framework to test
  algorithms of similarity hashing}}, \bibinfo{journal}{Digital Investigation}
  \bibinfo{volume}{10} (\bibinfo{year}{2013}{\natexlab{b}})
  \bibinfo{pages}{S50--S58}, ISSN \bibinfo{issn}{17422876},
  \doi{\bibinfo{doi}{10.1016/j.diin.2013.06.006}},
  \urlprefix\url{http://linkinghub.elsevier.com/retrieve/pii/S1742287613000522}.

\bibitem[{Li et~al.(2004)Li, Chen, Li, Ma, and Vitanyi}]{Li2004}
\bibinfo{author}{M.~Li}, \bibinfo{author}{X.~Chen}, \bibinfo{author}{X.~Li},
  \bibinfo{author}{B.~Ma}, \bibinfo{author}{P.~M. Vitanyi},
  \bibinfo{title}{{The Similarity Metric}}, \bibinfo{journal}{IEEE Transactions
  on Information Theory} \bibinfo{volume}{50}~(\bibinfo{number}{12})
  (\bibinfo{year}{2004}) \bibinfo{pages}{3250--3264}, ISSN
  \bibinfo{issn}{0018-9448}, \doi{\bibinfo{doi}{10.1109/TIT.2004.838101}}.

\bibitem[{Cebri{\'{a}}n et~al.(2005)Cebri{\'{a}}n, Alfonseca, Ortega, and
  {others}}]{cebrian2005common}
\bibinfo{author}{M.~Cebri{\'{a}}n}, \bibinfo{author}{M.~Alfonseca},
  \bibinfo{author}{A.~Ortega}, \bibinfo{author}{{others}},
  \bibinfo{title}{{Common pitfalls using the normalized compression distance:
  What to watch out for in a compressor}}, \bibinfo{journal}{Communications in
  Information {\&} Systems} \bibinfo{volume}{5}~(\bibinfo{number}{4})
  (\bibinfo{year}{2005}) \bibinfo{pages}{367--384}.

\bibitem[{Cebrin et~al.(2007)Cebrin, Alfonseca, and Ortega}]{Cebrin2007}
\bibinfo{author}{M.~Cebrin}, \bibinfo{author}{M.~Alfonseca},
  \bibinfo{author}{A.~Ortega}, \bibinfo{title}{{The Normalized Compression
  Distance Is Resistant to Noise}}, \bibinfo{journal}{IEEE Transactions on
  Information Theory} \bibinfo{volume}{53}~(\bibinfo{number}{5})
  (\bibinfo{year}{2007}) \bibinfo{pages}{1895--1900}, ISSN
  \bibinfo{issn}{0018-9448}, \doi{\bibinfo{doi}{10.1109/TIT.2007.894669}},
  \urlprefix\url{http://ieeexplore.ieee.org/document/4167725/}.

\bibitem[{Borbely(2015)}]{Borbely2015}
\bibinfo{author}{R.~S. Borbely}, \bibinfo{title}{{On normalized compression
  distance and large malware}}, \bibinfo{journal}{Journal of Computer Virology
  and Hacking Techniques}  (\bibinfo{year}{2015}) \bibinfo{pages}{1--8}ISSN
  \bibinfo{issn}{2263-8733}, \doi{\bibinfo{doi}{10.1007/s11416-015-0260-0}}.

\bibitem[{Pavlov(2007)}]{lzma07url}
\bibinfo{author}{I.~Pavlov}, \bibinfo{title}{{LZMA SDK (Software Development
  Kit)}}, \bibinfo{publisher}{http://www.7-zip.org/sdk.html},
  \urlprefix\url{http://www.7-zip.org/sdk.html}, \bibinfo{year}{2007}.

\bibitem[{Alshahwan et~al.(2015)Alshahwan, Barr, Clark, and
  Danezis}]{Alshahwan2015}
\bibinfo{author}{N.~Alshahwan}, \bibinfo{author}{E.~T. Barr},
  \bibinfo{author}{D.~Clark}, \bibinfo{author}{G.~Danezis},
  \bibinfo{title}{{Detecting Malware with Information Complexity}}
  \urlprefix\url{http://arxiv.org/abs/1502.07661}.

\bibitem[{Ziv and Lempel(1977)}]{Ziv1977}
\bibinfo{author}{J.~Ziv}, \bibinfo{author}{A.~Lempel}, \bibinfo{title}{{A
  universal algorithm for sequential data compression}}, \bibinfo{journal}{IEEE
  Transactions on Information Theory}
  \bibinfo{volume}{23}~(\bibinfo{number}{3}) (\bibinfo{year}{1977})
  \bibinfo{pages}{337--343}, ISSN \bibinfo{issn}{0018-9448},
  \doi{\bibinfo{doi}{10.1109/TIT.1977.1055714}},
  \urlprefix\url{http://ieeexplore.ieee.org/document/1055714/}.

\bibitem[{Broder(1997)}]{Broder:1997:RCD:829502.830043}
\bibinfo{author}{A.~Z. Broder}, \bibinfo{title}{{On the Resemblance and
  Containment of Documents}}, in: \bibinfo{booktitle}{Proceedings of the
  Compression and Complexity of Sequences 1997}, SEQUENCES '97,
  \bibinfo{publisher}{IEEE Computer Society}, \bibinfo{address}{Washington, DC,
  USA}, ISBN \bibinfo{isbn}{0-8186-8132-2}, \bibinfo{pages}{21--29},
  \urlprefix\url{http://dl.acm.org/citation.cfm?id=829502.830043},
  \bibinfo{year}{1997}.

\bibitem[{Broder et~al.(1998)Broder, Charikar, Frieze, and
  Mitzenmacher}]{Broder:1998:MIP:276698.276781}
\bibinfo{author}{A.~Z. Broder}, \bibinfo{author}{M.~Charikar},
  \bibinfo{author}{A.~M. Frieze}, \bibinfo{author}{M.~Mitzenmacher},
  \bibinfo{title}{{Min-wise Independent Permutations (Extended Abstract)}}, in:
  \bibinfo{booktitle}{Proceedings of the Thirtieth Annual ACM Symposium on
  Theory of Computing}, STOC '98, \bibinfo{publisher}{ACM},
  \bibinfo{address}{New York, NY, USA}, ISBN \bibinfo{isbn}{0-89791-962-9},
  \bibinfo{pages}{327--336}, \doi{\bibinfo{doi}{10.1145/276698.276781}},
  \urlprefix\url{http://doi.acm.org/10.1145/276698.276781},
  \bibinfo{year}{1998}.

\bibitem[{Breitinger et~al.(2012)Breitinger, Baier, and
  Beckingham}]{Breitinger2012}
\bibinfo{author}{F.~Breitinger}, \bibinfo{author}{H.~Baier},
  \bibinfo{author}{J.~Beckingham}, \bibinfo{title}{{Security and implementation
  analysis of the similarity digest sdhash}}, in: \bibinfo{booktitle}{First
  International Baltic Conference on Network Security {\&} Forensics (NeSeFo)},
  \bibinfo{year}{2012}.

\bibitem[{Levenshtein(1966)}]{levenshtein1966binary}
\bibinfo{author}{V.~I. Levenshtein}, \bibinfo{title}{{Binary codes capable of
  correcting deletions, insertions and reversals}}, in:
  \bibinfo{booktitle}{Soviet physics doklady}, vol.~\bibinfo{volume}{10},
  \bibinfo{pages}{707}, \bibinfo{year}{1966}.

\bibitem[{Knuth(1998)}]{Knuth:1998:ACP:280635}
\bibinfo{author}{D.~E. Knuth}, \bibinfo{title}{{The Art of Computer
  Programming, Volume 3: (2Nd Ed.) Sorting and Searching}},
  \bibinfo{publisher}{Addison Wesley Longman Publishing Co., Inc.},
  \bibinfo{address}{Redwood City, CA, USA}, ISBN \bibinfo{isbn}{0-201-89685-0},
  \bibinfo{year}{1998}.

\bibitem[{Dor and Zwick(2001)}]{Dor:2001:MSR:587900.587932}
\bibinfo{author}{D.~Dor}, \bibinfo{author}{U.~Zwick}, \bibinfo{title}{{Median
  Selection Requires (2+{$\epsilon$})N Comparisons}}, \bibinfo{journal}{SIAM J.
  Discret. Math.} \bibinfo{volume}{14}~(\bibinfo{number}{3})
  (\bibinfo{year}{2001}) \bibinfo{pages}{312--325}, ISSN
  \bibinfo{issn}{0895-4801}, \doi{\bibinfo{doi}{10.1137/S0895480199353895}},
  \urlprefix\url{http://dx.doi.org/10.1137/S0895480199353895}.

\bibitem[{Wong and Stamp(2006)}]{Wong2006}
\bibinfo{author}{W.~Wong}, \bibinfo{author}{M.~Stamp}, \bibinfo{title}{{Hunting
  for metamorphic engines}}, \bibinfo{journal}{Journal in Computer Virology}
  \bibinfo{volume}{2}~(\bibinfo{number}{3}) (\bibinfo{year}{2006})
  \bibinfo{pages}{211--229}, ISSN \bibinfo{issn}{1772-9904},
  \doi{\bibinfo{doi}{10.1007/s11416-006-0028-7}},
  \urlprefix\url{http://dx.doi.org/10.1007/s11416-006-0028-7}.

\bibitem[{Konstantinou(2008)}]{RHUL-MA-2008-02}
\bibinfo{author}{E.~Konstantinou}, \bibinfo{title}{{Metamorphic Virus: Analysis
  and Detection}}, \bibinfo{type}{Tech. Rep.}, \bibinfo{institution}{Royal
  Holloway University of London},
  \urlprefix\url{http://digirep.rhul.ac.uk/items/bde3a9fe-51c0-a19a-e04d-b324c0926a4a/1/},
  \bibinfo{year}{2008}.

\bibitem[{mic(2015)}]{microsoft_kaggle_2015}
\bibinfo{title}{{Microsoft Malware Classification Challenge (BIG 2015)}},
  \urlprefix\url{https://www.kaggle.com/c/malware-classification/},
  \bibinfo{year}{2015}.

\bibitem[{Arp et~al.(2014)Arp, Spreitzenbarth, Malte, Gascon, and
  Rieck}]{Arp2014}
\bibinfo{author}{D.~Arp}, \bibinfo{author}{M.~Spreitzenbarth},
  \bibinfo{author}{H.~Malte}, \bibinfo{author}{H.~Gascon},
  \bibinfo{author}{K.~Rieck}, \bibinfo{title}{{Drebin: Effective and
  Explainable Detection of Android Malware in Your Pocket}},
  \bibinfo{journal}{Symposium on Network and Distributed System Security
  (NDSS)} ~(\bibinfo{number}{February}) (\bibinfo{year}{2014})
  \bibinfo{pages}{23--26}, \doi{\bibinfo{doi}{10.14722/ndss.2014.23247}}.

\bibitem[{Brodersen et~al.(2010)Brodersen, Ong, Stephan, and
  Buhmann}]{Brodersen:2010:BAP:1904935.1905533}
\bibinfo{author}{K.~H. Brodersen}, \bibinfo{author}{C.~S. Ong},
  \bibinfo{author}{K.~E. Stephan}, \bibinfo{author}{J.~M. Buhmann},
  \bibinfo{title}{{The Balanced Accuracy and Its Posterior Distribution}}, in:
  \bibinfo{booktitle}{Proceedings of the 2010 20th International Conference on
  Pattern Recognition}, ICPR '10, \bibinfo{publisher}{IEEE Computer Society},
  \bibinfo{address}{Washington, DC, USA}, ISBN
  \bibinfo{isbn}{978-0-7695-4109-9}, \bibinfo{pages}{3121--3124},
  \doi{\bibinfo{doi}{10.1109/ICPR.2010.764}},
  \urlprefix\url{http://dx.doi.org/10.1109/ICPR.2010.764},
  \bibinfo{year}{2010}.

\bibitem[{Roussev(2009)}]{2009:BBS:1488734.1490103}
\bibinfo{author}{V.~Roussev}, \bibinfo{title}{{Building a Better Similarity
  Trap with Statistically Improbable Features}}, in:
  \bibinfo{booktitle}{Proceedings of the 42Nd Hawaii International Conference
  on System Sciences}, HICSS '09, \bibinfo{publisher}{IEEE Computer Society},
  \bibinfo{address}{Washington, DC, USA}, ISBN
  \bibinfo{isbn}{978-0-7695-3450-3}, \bibinfo{pages}{1--10},
  \doi{\bibinfo{doi}{10.1109/HICSS.2009.97}},
  \urlprefix\url{http://dx.doi.org/10.1109/HICSS.2009.97},
  \bibinfo{year}{2009}.

\bibitem[{Winter et~al.(2013{\natexlab{b}})Winter, Schneider, and
  Yannikos}]{Winter2013a}
\bibinfo{author}{C.~Winter}, \bibinfo{author}{M.~Schneider},
  \bibinfo{author}{Y.~Yannikos}, \bibinfo{title}{{F2S2: Fast forensic
  similarity search through indexing piecewise hash signatures}},
  \bibinfo{journal}{Digital Investigation}
  \bibinfo{volume}{10}~(\bibinfo{number}{4})
  (\bibinfo{year}{2013}{\natexlab{b}}) \bibinfo{pages}{361--371}, ISSN
  \bibinfo{issn}{17422876}, \doi{\bibinfo{doi}{10.1016/j.diin.2013.08.003}},
  \urlprefix\url{http://linkinghub.elsevier.com/retrieve/pii/S1742287613000789}.

\bibitem[{Breitinger et~al.(2014{\natexlab{a}})Breitinger, Baier, and
  White}]{Breitinger2014}
\bibinfo{author}{F.~Breitinger}, \bibinfo{author}{H.~Baier},
  \bibinfo{author}{D.~White}, \bibinfo{title}{{On the database lookup problem
  of approximate matching}}, \bibinfo{journal}{Digital Investigation}
  \bibinfo{volume}{11} (\bibinfo{year}{2014}{\natexlab{a}})
  \bibinfo{pages}{S1--S9}, ISSN \bibinfo{issn}{17422876},
  \doi{\bibinfo{doi}{10.1016/j.diin.2014.03.001}},
  \urlprefix\url{http://linkinghub.elsevier.com/retrieve/pii/S1742287614000061}.

\bibitem[{Breitinger et~al.(2014{\natexlab{b}})Breitinger, Rathgeb, and
  Baier}]{DBLP:journals/jdfsl/BreitingerRB14}
\bibinfo{author}{F.~Breitinger}, \bibinfo{author}{C.~Rathgeb},
  \bibinfo{author}{H.~Baier}, \bibinfo{title}{{An Efficient Similarity Digests
  Database Lookup - A Logarithmic Divide {\&} Conquer Approach}},
  \bibinfo{journal}{The Journal of Digital Forensics, Security and Law (JDFSL)}
  \bibinfo{volume}{9}~(\bibinfo{number}{2})
  (\bibinfo{year}{2014}{\natexlab{b}}) \bibinfo{pages}{155--166},
  \urlprefix\url{http://ojs.jdfsl.org/index.php/jdfsl/article/view/276}.

\bibitem[{Lillis et~al.(2017)Lillis, Breitinger, and Scanlon}]{Lillis2017}
\bibinfo{author}{D.~Lillis}, \bibinfo{author}{F.~Breitinger},
  \bibinfo{author}{M.~Scanlon}, \bibinfo{title}{{Expediting MRSH-v2 Approximate
  Matching with Hierarchical Bloom Filter Trees}}, in: \bibinfo{booktitle}{9th
  EAI International Conference on Digital Forensics and Cyber Crime (ICDF2C
  2017)}, \bibinfo{publisher}{Springer}, \bibinfo{address}{Prague, Czechia},
  \bibinfo{year}{2017}.

\bibitem[{Yianilos(1993)}]{Yianilos1993}
\bibinfo{author}{P.~Yianilos}, \bibinfo{title}{{Data structures and algorithms
  for nearest neighbor search in general metric spaces}}, in:
  \bibinfo{booktitle}{Proceedings of the fourth annual ACM-SIAM Symposium on
  Discrete algorithms}, \bibinfo{publisher}{Society for Industrial and Applied
  Mathematics}, \bibinfo{pages}{311–321},
  \urlprefix\url{http://dl.acm.org/citation.cfm?id=313789},
  \bibinfo{year}{1993}.

\bibitem[{Uhlmann(1991)}]{Uhlmann1991}
\bibinfo{author}{J.~K. Uhlmann}, \bibinfo{title}{{Satisfying general proximity
  / similarity queries with metric trees}}, \bibinfo{journal}{Information
  Processing Letters} \bibinfo{volume}{40}~(\bibinfo{number}{4})
  (\bibinfo{year}{1991}) \bibinfo{pages}{175--179}, ISSN
  \bibinfo{issn}{00200190}, \doi{\bibinfo{doi}{10.1016/0020-0190(91)90074-R}},
  \urlprefix\url{http://linkinghub.elsevier.com/retrieve/pii/002001909190074R}.

\bibitem[{Beygelzimer et~al.(2006)Beygelzimer, Kakade, and
  Langford}]{Beygelzimer2006}
\bibinfo{author}{A.~Beygelzimer}, \bibinfo{author}{S.~Kakade},
  \bibinfo{author}{J.~Langford}, \bibinfo{title}{{Cover trees for nearest
  neighbor}}, in: \bibinfo{booktitle}{International Conference on Machine
  Learning}, \bibinfo{publisher}{ACM}, \bibinfo{address}{New York},
  \bibinfo{pages}{97--104},
  \urlprefix\url{http://www.cs.princeton.edu/courses/archive/spr05/cos598E/bib/covertree.pdf},
  \bibinfo{year}{2006}.

\bibitem[{Izbicki and Shelton(2015)}]{Izbicki2015}
\bibinfo{author}{M.~Izbicki}, \bibinfo{author}{C.~R. Shelton},
  \bibinfo{title}{{Faster Cover Trees}}, in: \bibinfo{booktitle}{Proceedings of
  the Thirty-Second International Conference on Machine Learning},
  vol.~\bibinfo{volume}{37}, \bibinfo{year}{2015}.

\bibitem[{Baier and Breitinger(2011)}]{Baier2011}
\bibinfo{author}{H.~Baier}, \bibinfo{author}{F.~Breitinger},
  \bibinfo{title}{{Security Aspects of Piecewise Hashing in Computer
  Forensics}}, in: \bibinfo{booktitle}{2011 Sixth International Conference on
  IT Security Incident Management and IT Forensics}, \bibinfo{publisher}{IEEE},
  ISBN \bibinfo{isbn}{978-1-4577-0146-7}, \bibinfo{pages}{21--36},
  \doi{\bibinfo{doi}{10.1109/IMF.2011.16}},
  \urlprefix\url{http://ieeexplore.ieee.org/document/5931110/},
  \bibinfo{year}{2011}.

\end{thebibliography}

\appendix
\section{Full FRASH Results} \label{sec:full_frash}

Here in the appendix we provide more complete results from the FRASH tests for those who are interested. In these full tables, \textit{Score} is the average score for each match, and \textit{Matches} or \textit{Match} it the absolute number of matches at that size.

\FloatBarrier
\subsection{Single Common Block Tables}

In these tables, we show the average block size percentage as the \textit{Size} column. The associated average block size can be computed from these tables by multiplying the total block size of the table, with the percentage given in each column. The SCB tests were run for 50 trials each. This covers the results in \autoref{tbl:scb_full_512kb}, \autoref{tbl:scb_full_2mb}, and \autoref{tbl:scb_full_8mb}. 

\begin{table}[!htbp]
\centering
\caption{Complete Single Common Block results for a total block size of 512 KB.}
\label{tbl:scb_full_512kb}
\begin{adjustbox}{max width=\linewidth}
\begin{tabular}{@{}ccccccc@{}}
\toprule
               & \multicolumn{2}{c}{ssdeep} & \multicolumn{2}{c}{sdhash} & \multicolumn{2}{c}{LZJD}  \\
\cmidrule(lr){2-3} \cmidrule(lr){4-5} \cmidrule(lr){6-7}
Score           & Size (\%) & Matches      &  Size (\%) & Matches      & Size (\%)  & Matches      \\ 
\midrule
  $\geq$65  & 48.44 & 2 &       ---  &       ---  &       ---  &       --- \\
  $\geq$60  & 47.54 & 14 &       ---  &       ---  &       ---  &       --- \\
  $\geq$55  & 43.59 & 20 &       ---  &       ---  &       ---  &       --- \\
  $\geq$50  & 42.63 & 39 &       ---  &       ---  &       ---  &       --- \\
  $\geq$45  & 38.78 & 44 & 48.96 & 6 &       ---  &       --- \\
  $\geq$40  & 32.42 & 48 & 46.61 & 12 &       ---  &       --- \\
  $\geq$35  & 26.45 & 43 & 42.38 & 16 &       ---  &       --- \\
  $\geq$30  & 22.94 & 41 & 40.85 & 28 &       ---  &       --- \\
   $\geq$25  & 18.42 & 28 & 36.13 & 32 & 44.15 & 39\\
   $\geq$20  & 20.62 & 5 & 31.33 & 37 & 15.75 & 50\\
   $\geq$15  &        ---  &        ---  & 27.27 & 44 & 3.25 & 50\\
  $\geq$10  &       ---  &       ---  & 20.88 & 50 &       ---  &       --- \\
    $\geq$5  &        ---  &        ---  & 10.88 & 50 &        ---  &        --- \\
0 & 15.5 & 50 & 3.12 & 50 &        ---  &        --- \\
\bottomrule
\end{tabular}
\end{adjustbox}
\end{table}

\begin{table}[!htbp]
\centering
\caption{Complete Single Common Block results for a total block size of 2 MB.}
\label{tbl:scb_full_2mb}
\begin{adjustbox}{max width=\linewidth}
\begin{tabular}{@{}ccccccc@{}}
\toprule
               & \multicolumn{2}{c}{ssdeep} & \multicolumn{2}{c}{sdhash} & \multicolumn{2}{c}{LZJD}  \\
\cmidrule(lr){2-3} \cmidrule(lr){4-5} \cmidrule(lr){6-7}
Score           &  Size (\%) & Matches      &   Size (\%) & Matches      &  Size (\%)  & Matches      \\ 
\midrule
  $\geq$70  & 44.92 & 2 &       ---  &       ---  &       ---  &       --- \\
  $\geq$65  & 44.82 & 8 &       ---  &       ---  &       ---  &       --- \\
  $\geq$60  & 43.36 & 12 &       ---  &       ---  &       ---  &       --- \\
  $\geq$55  & 42.86 & 28 &       ---  &       ---  &       ---  &       --- \\
  $\geq$50  & 40.28 & 41 &       ---  &       ---  &       ---  &       --- \\
  $\geq$45  & 37.27 & 48 & 47.27 & 4 &       ---  &       --- \\
  $\geq$40  & 30.05 & 49 & 46.03 & 12 &       ---  &       --- \\
  $\geq$35  & 25.98 & 48 & 42.68 & 19 &       ---  &       --- \\
  $\geq$30  & 21.54 & 46 & 39.84 & 27 &       ---  &       --- \\
  $\geq$25  & 18.85 & 23 & 35.66 & 34 &       ---  &       --- \\
  $\geq$20  & 19.98 & 7 & 31.38 & 42 &       ---  &       --- \\
  $\geq$15  &       ---  &       ---  & 24.47 & 44 & 42.39 & 46\\
  $\geq$10  &       ---  &       ---  & 18.72 & 50 & 18.38 & 50\\
   $\geq$5  &       ---  &       ---  & 9.17 & 50 & 0.78 & 50\\
0 & 19.17 & 50 & 0.88 & 50 &        ---  &        --- \\
\bottomrule
\end{tabular}
\end{adjustbox}
\end{table}

\begin{table}[!htbp]
\centering
\caption{Complete Single Common Block results for a total block size of 8 MB.}
\label{tbl:scb_full_8mb}
\begin{adjustbox}{max width=\linewidth}
\begin{tabular}{@{}ccccccc@{}}
\toprule
               & \multicolumn{2}{c}{ssdeep} & \multicolumn{2}{c}{sdhash} & \multicolumn{2}{c}{LZJD}  \\
\cmidrule(lr){2-3} \cmidrule(lr){4-5} \cmidrule(lr){6-7}
Score           &  Size (\%) & Matches      &   Size (\%) & Matches      &  Size (\%)  & Matches      \\ 
\midrule
   $\geq$70  & 44.92 & 1 &        ---  &        ---  &        ---  &        --- \\
   $\geq$65  & 46.7 & 8 &        ---  &        ---  &        ---  &        --- \\
   $\geq$60  & 45.81 & 20 &        ---  &        ---  &        ---  &        --- \\
  $\geq$55  & 42.55 & 31 &       ---  &       ---  &       ---  &       --- \\
   $\geq$50  & 39.05 & 40 & 49.9 & 2 &        ---  &        --- \\
   $\geq$45  & 35.83 & 48 & 47.01 & 7 &        ---  &        --- \\
   $\geq$40  & 28.52 & 50 & 43.86 & 13 &        ---  &        --- \\
   $\geq$35  & 23.74 & 49 & 41.94 & 22 &        ---  &        --- \\
   $\geq$30  & 20.07 & 50 & 37.3 & 25 &        ---  &        --- \\
   $\geq$25  & 18.46 & 36 & 33.41 & 31 &        ---  &        --- \\
   $\geq$20  & 16.41 & 9 & 28.36 & 35 & 49.02 & 1\\
   $\geq$15  & 12.89 & 1 & 26.47 & 44 & 36.56 & 50\\
   $\geq$10  &        ---  &        ---  & 19.75 & 50 & 6.8 & 50\\
   $\geq$5  &       ---  &       ---  & 9.51 & 50 & 0.2 & 50\\
0 & 17.61 & 50 & 0.99 & 50 &        ---  &        --- \\
\bottomrule
\end{tabular}
\end{adjustbox}
\end{table}

\FloatBarrier
\subsection{Random Noise Table}

The full results from the random noise test are given in \autoref{tbl:scb_obfuscation_full}. The \textit{Change} column is the average percent of bytes in the filed that needed to be edited for a score of that value to be obtained, and \textit{Match} is the number of files that FRASH was able to successfully reduce to the given score range. The most robust method for each score is shown in \textbf{bold}. The default spacing used in FRASH is 10, but we reduced the spacing to 5 to take advantage of LZJD's performance of LZJD. The high resitance of LZJD meant that a zero value was never produced, which did not interact well with FRASH's execution. The second to last row shows that for 7 of the 100 files, a match score in the range of $[1, 5)$ was produced by modifying an average of 32\% of the file. This value is artificially low, as almost all tests were stopped prematurely before LZJD even reached a score of 15. 

\begin{table}[!htbp]
\centering
\caption{Random Noise Test. }
\label{tbl:scb_obfuscation_full}
\begin{adjustbox}{max width=\linewidth}
\begin{tabular}{@{}ccccccc@{}}
\toprule
               & \multicolumn{2}{c}{ssdeep} & \multicolumn{2}{c}{sdhash} & \multicolumn{2}{c}{LZJD}  \\
\cmidrule(lr){2-3} \cmidrule(lr){4-5} \cmidrule(lr){6-7}
Score           &  Change (\%) & Match      &   Change (\%) & Match      &  Change (\%)  & Match      \\ 
\midrule
$\geq$95  & 0.00058 & 100 & \textbf{0.01293} & 84 & 0.00151 & 60\\
$\geq$90  & 0.00136 & 99 & \textbf{0.02655} & 90 & 0.00317 & 46\\
$\geq$85  & 0.00211 & 99 & \textbf{0.04710} & 91 & 0.00490 & 55\\
$\geq$80  & 0.00291 & 99 & \textbf{0.06560} & 93 & 0.00777 & 59\\
$\geq$75  & 0.00398 & 93 & \textbf{0.09060} & 95 & 0.01243 & 65\\
$\geq$70  & 0.00524 & 88 & \textbf{0.10677} & 96 & 0.02378 & 75\\
$\geq$65  & 0.00814 & 73 & \textbf{0.13483} & 97 & 0.07260 & 86\\
$\geq$60  & 0.01214 & 51 & \textbf{0.17455} & 97 & 0.16215 & 95\\
$\geq$55  & 0.01676 & 30 & 0.22571 & 96 & \textbf{0.36321} & 95\\
$\geq$50  & 0.02057 & 18 & 0.27750 & 99 & \textbf{0.60610} & 96\\
$\geq$45  & 0.04409 & 10 & 0.32868 & 100 & \textbf{1.14643} & 99\\
$\geq$40  & 0.06148 & 7 & 0.39398 & 100 & \textbf{1.96722} & 99\\
$\geq$35  & 0.04466 & 3 & 0.45643 & 100 & \textbf{3.44188} & 100\\
$\geq$30  & 0.05740 & 2 & 0.54923 & 100 & \textbf{6.22476} & 100\\
$\geq$25  &  ---  &  ---  & 0.62586 & 99 & \textbf{10.99275} & 100\\
$\geq$20  &  ---  &  ---  & 0.69417 & 99 & \textbf{18.03654} & 100\\
$\geq$15  &  ---  &  ---  & 0.84723 & 98 & \textbf{30.56353} & 90\\
$\geq$10  &  ---  &  ---  & 0.97390 & 99 & \textbf{48.63043} & 77\\
$\geq$5  &  ---  &  ---  & 1.14414 & 100 & \textbf{63.80973} & 57\\
$[1, 5)$ & --- & --- & --- & --- & 32.03773 & 7\\
0 & 0.01283 & 100 & 1.55763 & 100 & --- & ---\\
\bottomrule
\end{tabular}
\end{adjustbox}
\end{table}

\begin{figure}
\centering
\begin{tikzpicture}
\begin{axis}[
	legend pos=outer south,
    legend columns=3, 
    xlabel=Percentage of File Changed,
    ylabel=Number Files Matched,
    xmode=log,
    ]
    \addplot[dashed,green,mark=square,mark options={solid}] table [x=ssdeep_change, y=ssdeep_matches, col sep=comma] {csvs/random_noise.csv};
    \addplot[dashed,blue,mark=o,mark options={solid}] table [x=sdhash_change, y=sdhash_matches, col sep=comma] {csvs/random_noise.csv};
    \addplot[red,mark=x] table [x=lzjd_change, y=lzjd_matches, col sep=comma] {csvs/random_noise.csv};
    \legend{ssdeep,sdhash,LZJD}

\end{axis}
\end{tikzpicture}
\caption{Random Noise results, plotted for each method showing how much of the file can be changed while still obtaining a correct match. }
\label{fig:rand_noise}
\end{figure}
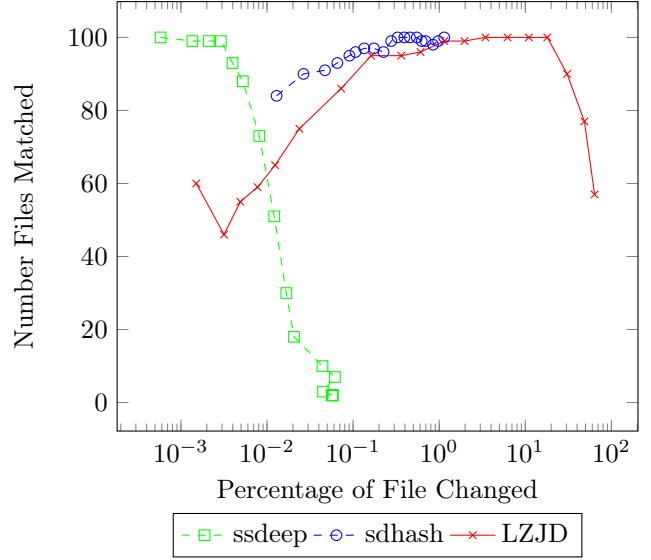

The change and match values in this table are also shown in \autoref{fig:rand_noise}, which plots the number of files matched against the the percentage of the file changed. Note that the x-axis is on a log-scale. This score can go up and down because it is based on the number of files matched receiving a minimum score (i.e., a score $\geq 90$). Because of this interpretation of the figure must be done carefully, and emphasize that this figure is to demonstrate the range of byte alterations each method can withstand. In this light is becomes clear that ssdeep is only able to produce matches when very little of the file has been altered, less than 0.1\%. Sdhash is able to perform matches in a range up to 1.1\% of the file being randomly altered, but fails to produce any matches past this point. LZJD in contrast is able to suffer from as much as 63\% of the file being randomly altered, and still over half the files. It is the only method to cover this large of a range in the amount of bytes that can be altered. Again, we note that the lower number of matches obtained by LZJD and sdhash in the left-most portion of the plot are because the associated minimum score is not factored in. For example, the 60 matches of LZJD at 0.002\% is not indicating that only 60 of the files could be matched after that percentage of files changed. It's value is 60 because only 60 files could be matched \textit{and} obtain a similarity score $\geq 95$. 

\FloatBarrier
\subsection{Fragment Test Tables}

Tables \autoref{tbl:fragment_random_full} and \autoref{tbl:fragment_right_full} are the complete version of \autoref{tbl:frag_randStart} and \autoref{tbl:frag_right} respectively. The \textit{Size} column is the percent file size. We can see that when LZJD and Sdhash don't get every match, LZJD always has more matches. It is also more clear that Sdhash's performance degrades at around 5\% and drops quickly, where LZJD is more robust in being able to still hit matches. 

\begin{table}[!htbp]
\centering
\caption{Fragment detection test result (cut side: random start, then alternating). Matches gives the }
\label{tbl:fragment_random_full}
\begin{adjustbox}{max width=\linewidth}
\begin{tabular}{@{}ccccccc@{}}
\toprule
               & \multicolumn{2}{c}{ssdeep} & \multicolumn{2}{c}{sdhash} & \multicolumn{2}{c}{LZJD}  \\
\cmidrule(lr){2-3} \cmidrule(lr){4-5} \cmidrule(lr){6-7}
Size           & Score       & Matches      &      Score  & Matches      & Score      & Matches      \\ 
\midrule
95             & 96.7        & 4454         & 83.4        & \textbf{4457}         & 72.4       & \textbf{4457}         \\
90             & 92.3        & 4452         & 70.1        & \textbf{4457}         & 49.2       & \textbf{4457}         \\
85             & 89.5        & 4442         & 69.9        & \textbf{4457}         & 45.5       & \textbf{4457}         \\
80             & 86.5        & 4433         & 69.1        & \textbf{4457}         & 41.9       & \textbf{4457}         \\
75             & 83.5        & 4417         & 68.3        & \textbf{4457}         & 38.8       & \textbf{4457}         \\
70             & 80.1        & 4403         & 68.0        & \textbf{4457}         & 35.7       & \textbf{4457}         \\
65             & 76.7        & 4367         & 68.4        & \textbf{4457}         & 32.9       & \textbf{4457}         \\
60             & 73.2        & 4321         & 68.4        & \textbf{4457}         & 30.1       & \textbf{4457}         \\
55             & 69.7        & 4205         & 68.0        & \textbf{4457}         & 27.5       & \textbf{4457}         \\
50             & 65.9        & 4071         & 68.5        & \textbf{4457}         & 24.9       & \textbf{4457}         \\
45             & 62.4        & 3699         & 69.2        & \textbf{4457}         & 22.4       & \textbf{4457}         \\
40             & 58.6        & 3140         & 69.8        & \textbf{4457}         & 20.1       & \textbf{4457}         \\
35             & 54.7        & 2477         & 71.0        & \textbf{4457}         & 17.8       & \textbf{4457}         \\
30             & 51.2        & 1704         & 71.4        & \textbf{4457}         & 15.5       & \textbf{4457}         \\
25             & 47.9        & 928          & 72.2        & \textbf{4456}         & 13.2       & \textbf{4456}         \\
20             & 45.7        & 411          & 73.1        & 4453         & 11.0       & \textbf{4456}         \\
15             & 43.7        & 132          & 73.9        & 4450         & 8.8        & \textbf{4456}         \\
10             & 46.2        & 29           & 75.7        & 4371         & 6.4        & \textbf{4456}         \\
5              & 61.0        & 2            & 77.4        & 4036         & 3.9        & \textbf{4454}         \\
4              & ---         & ---          & 78.4        & 3838         & 3.3        & \textbf{4444}         \\
3              & ---         & ---          & 78.7        & 3616         & 2.7        & \textbf{4432}         \\
2              & ---         & ---          & 79.1        & 3257         & 2.0        & \textbf{4419}         \\
1              & ---         & ---          & 81.0        & 2581         & 1.3        & \textbf{4390}        \\ \bottomrule
\end{tabular}
\end{adjustbox}
\end{table}

\begin{table}[!htbp]
\centering
\caption{Fragment detection test result Fragment detection test result (cut side: right (end only), 5 \%). Matches gives the }
\label{tbl:fragment_right_full}
\begin{adjustbox}{max width=\linewidth}
\begin{tabular}{@{}ccccccc@{}}
\toprule
               & \multicolumn{2}{c}{ssdeep} & \multicolumn{2}{c}{sdhash} & \multicolumn{2}{c}{LZJD}  \\
\cmidrule(lr){2-3} \cmidrule(lr){4-5} \cmidrule(lr){6-7}
Size           & Score       & Matches      &      Score  & Matches      & Score      & Matches      \\ 
\midrule
95	&	97.70	&	\textbf{4457}	&	97.28	&	\textbf{4457}	&	92.80	&	\textbf{4457}  \\
90	&	95.90	&	4456	&	98.17	&	\textbf{4457}	&	85.96	&	\textbf{4457}  \\
85	&	93.81	&	4453	&	98.86	&	\textbf{4457}	&	79.33	&	\textbf{4457}  \\
80	&	91.42	&	4444	&	99.32	&	\textbf{4457}	&	72.99	&	\textbf{4457}  \\
75	&	88.85	&	4440	&	99.37	&	\textbf{4457}	&	66.87	&	\textbf{4457}  \\
70	&	85.92	&	4429	&	99.44	&	\textbf{4457}	&	61.01	&	\textbf{4457}  \\
65	&	82.79	&	4414	&	99.49	&	\textbf{4457}	&	55.39	&	\textbf{4457}  \\
60	&	79.36	&	4378	&	99.46	&	\textbf{4457}	&	50.08	&	\textbf{4457}  \\
55	&	75.68	&	4307	&	99.49	&	\textbf{4457}	&	44.98	&	\textbf{4457}  \\
50	&	71.73	&	4148	&	99.51	&	\textbf{4457}	&	40.06	&	\textbf{4457}  \\
45	&	68.01	&	3815	&	99.41	&	\textbf{4457}	&	35.46	&	\textbf{4457}  \\
40	&	64.39	&	3326	&	99.37	&	\textbf{4457}	&	31.11	&	\textbf{4457}  \\
35	&	61.15	&	2697	&	99.28	&	\textbf{4457}	&	26.97	&	\textbf{4457}  \\
30	&	58.26	&	1968	&	99.20	&	\textbf{4457}	&	23.00	&	\textbf{4457}  \\
25	&	56.76	&	1191	&	99.07	&	\textbf{4457}	&	19.21	&	\textbf{4457}  \\
20	&	55.35	&	632	&	98.83	&	\textbf{4457}	&	15.54	&	\textbf{4457}  \\
15	&	53.21	&	289	&	98.46	&	\textbf{4457}	&	11.94	&	\textbf{4457}  \\
10	&	56.90	&	77	&	97.88	&	4380	&	8.33	&	\textbf{4457}  \\
5	&	55.68	&	22	&	96.93	&	4061	&	4.63	&	\textbf{4457}  \\
4	&	59.00	&	12	&	96.40	&	3896	&	3.88	&	\textbf{4457}  \\
3	&	47.89	&	9	&	95.04	&	3678	&	3.09	&	\textbf{4457}  \\
2	&	66.00	&	2	&	93.09	&	3304	&	2.23	&	\textbf{4456}  \\
1	&	     --- 	&	      ---	&	90.50	&	2617	&	1.36	&	\textbf{4442}  \\
\bottomrule
\end{tabular}
\end{adjustbox}
\end{table}

\FloatBarrier
\subsection{Alignment Table Results}

In \autoref{tbl:alignment_fixed_size_full} we provide the alignment results for adding $\leq$64 KB to the query file. This is the easier range of the test, and we can see that both sdhash and LZJD successfully match all files regardless of added bytes. 

\begin{table}[!htbp]
\centering
\caption{Alignment test result (fixed size, step size = 4 KB, max size = 64 KB) }
\label{tbl:alignment_fixed_size_full}
\begin{adjustbox}{max width=\linewidth}
\begin{tabular}{@{}ccccccc@{}}
\toprule
               & \multicolumn{2}{c}{ssdeep} & \multicolumn{2}{c}{sdhash} & \multicolumn{2}{c}{LZJD}  \\
\cmidrule(lr){2-3} \cmidrule(lr){4-5} \cmidrule(lr){6-7}
Added (KB)           & Score       & Matches      &      Score  & Matches      & Score      & Matches      \\ 
\midrule
   4  	&	91.31	&	4439	&	51.30	&	\textbf{4457}	&	42.38	&	\textbf{4457} \\
   8  	&	87.07	&	4279	&	78.51	&	\textbf{4457}	&	36.28	&	\textbf{4457} \\
  12  	&	84.39	&	4120	&	65.57	&	\textbf{4457}	&	32.60	&	\textbf{4457} \\
  16  	&	82.76	&	3901	&	64.22	&	\textbf{4457}	&	30.01	&	\textbf{4457} \\
  20  	&	82.19	&	3690	&	80.50	&	\textbf{4457}	&	28.04	&	\textbf{4457} \\
  24  	&	81.21	&	3580	&	51.58	&	\textbf{4457}	&	26.50	&	\textbf{4457} \\
  28  	&	79.98	&	3465	&	90.36	&	\textbf{4457}	&	25.20	&	\textbf{4457} \\
  32  	&	79.41	&	3314	&	52.38	&	\textbf{4457}	&	24.13	&	\textbf{4457} \\
  36  	&	79.49	&	3154	&	78.37	&	\textbf{4457}	&	23.21	&	\textbf{4457} \\
  40  	&	79.15	&	3059	&	65.83	&	\textbf{4457}	&	22.39	&	\textbf{4457} \\
  44  	&	79.34	&	2949	&	64.25	&	\textbf{4457}	&	21.65	&	\textbf{4457} \\
  48  	&	78.67	&	2895	&	80.62	&	\textbf{4457}	&	21.02	&	\textbf{4457} \\
  52  	&	78.03	&	2839	&	52.48	&	\textbf{4457}	&	20.40	&	\textbf{4457} \\
  56  	&	77.41	&	2775	&	88.19	&	\textbf{4457}	&	19.88	&	\textbf{4457} \\
  60  	&	76.65	&	2721	&	53.60	&	\textbf{4457}	&	19.39	&	\textbf{4457} \\
  64  	&	76.27	&	2645	&	78.05	&	\textbf{4457}	&	18.95	&	\textbf{4457} \\
\bottomrule
\end{tabular}
\end{adjustbox}
\end{table}

\end{document}